%% file: fh.tex
\documentstyle[12pt]{article}
\input{psbox}

\def\main{
    \pagestyle{headings}
    \renewcommand{\thefootnote}{\arabic{footnote}}
    \setcounter{footnote}{0}
    \setcounter{section}{0}
    \baselineskip = 0.6cm
    \pagestyle{plain}
    \renewcommand{\thesection}{\arabic{section}.}
    \renewcommand{\thesubsection}{\arabic{section}.\arabic{subsection}.}
    \renewcommand{\theequation}{\arabic{section}.\arabic{equation}}}
\def\appendix{
    \setcounter{section}{0}
    \renewcommand{\thesection}{Appendix \Alph{section}:}
    \renewcommand{\thesubsection}{\Alph{section}.\arabic{subsection}.}
    \renewcommand{\theequation}{\Alph{section}.\arabic{equation}}}
\def\sectiontitle#1{
    \vskip8mm\begin{center}{\bf\thesection~#1}\end{center}}
\def\subsectiontitle#1{
    \vskip6mm\noindent{\sc\thesubsection~#1}\vskip4mm}
\def\paragraphtitle#1{
    \vskip6mm\noindent{\it #1}\par}
\def\section#1{
    \addtocounter{section}{1}\setcounter{subsection}{0}
    \setcounter{equation}{0}\sectiontitle{#1}}
\def\subsection#1{
    \addtocounter{subsection}{1}\subsectiontitle{#1}}
\def\paragraph#1{
    \paragraphtitle{#1}}
\def\title#1{
    \baselineskip0.8cm\vskip2cm
    \begin{center}{\large\bf #1}\end{center}}
\def\author#1#2#3{
    \baselineskip0.6cm
    \begin{center}#1\footnote{\tt #2}\\\vskip2mm{\it #3}\vskip3mm
    \end{center}}
\def\abst#1{
    \vskip8mm\baselineskip=3.5ex
    \begin{center}{\bf Abstract}\end{center}\par\smallskip #1}

\def\thebibliography#1{\list
 {[\arabic{enumi}]}{\settowidth\labelwidth{[#1]}\leftmargin\labelwidth
  \advance\leftmargin\labelsep
  \usecounter{enumi}}
  \def\newblock{\hskip .11em plus .33em minus .07em}
  \sloppy\clubpenalty4000\widowpenalty4000
  \sfcode`\.=1000\relax}
\let\endthebibliography=\endlist

\newcommand{\bra}[1]{\left<{#1}\right|}
\newcommand{\ket}[1]{\left|{#1}\right>}
\newcommand{\vev}[1]{\left<{#1}\right>}
\newcommand{\normal}[1]{\;:\!{#1}\!:\;}
\newcommand{\tsum}{{\textstyle \sum}}
\newcommand{\ghyp}{{_3F_2}}
\newcommand{\sinpi}[1]{{\bf s}(#1)}
\newcommand{\fig}[2]{
   \begin{figure}[hbt]
   \begin{center}\mbox{\psbox{#1.ps}}\caption{#2}\label{#1}\end{center}
   \end{figure}}
\newcommand{\IIxI}[2]
{\left[\!\!\begin{array}{c}#1\\#2\end{array}\!\!\right]}
\newcommand{\IIxII}[4]
{\left[\!\!\begin{array}{cc}#1&#2\\#3&#4\end{array}\!\!\right]}

\begin{document}

\main

\thispagestyle{empty}
\setcounter{page}{0}

\baselineskip 5mm
\hfill\vbox{\hbox{YITP-01-43}
            \hbox{hep-th/0105217} }

\title{Three-point Functions in Sine-Liouville Theory}
\vskip10mm
\author{Takeshi Fukuda ~and~ Kazuo Hosomichi}
       {tfukuda@yukawa.kyoto-u.ac.jp,
        hosomiti@yukawa.kyoto-u.ac.jp}
       {Yukawa Institute for Theoretical Physics \\
        Kyoto University, Kyoto 606-8502, Japan}
\abst{
   We calculate the three-point functions in the
 sine-Liouville theory explicitly.
 The same calculation was done in the (unpublished) work
 of Fateev, Zamolodchikov and Zamolodchikov to check
 the conjectured duality between the sine-Liouville and the
 $SL(2,{\bf R})/U(1)$ coset CFTs.
 The evaluation of correlators boils down to that of
 a free-field theory with a certain number of insertion
 of screening operators.
 We prove that the winding number conservation is violated
 up to $\pm 1$ in three-point functions,
 which is in agreement with the result of FZZ
 that in generic $N$-point correlators
 the winding number conservation is violated up to $N-2$ units.
 A new integral formula of Dotsenko-Fateev type is derived,
 using which we write down the generic three-point functions
 of tachyons explicitly.
 When the winding number is conserved, the resultant expression
 is shown to reproduce the correlators in the coset model correctly,
 including the group-theoretical factor.
 As an application, we also study the superstring theory
 on linear dilaton background which is described by
 super-Liouville theory.
 We obtain the three-point amplitude of tachyons in which the winding
 number conservation is violated.
 }

\vspace*{\fill}
\noindent May~~2001
\newpage

\section{Introduction}

   The $SL(2,{\bf R})/U(1)$ coset model has a number of applications
 in recent topics in string theory.
 As was first studied in \cite{W,DVV},
 the model is known to describe a certain two-dimensional black hole,
 and serve as an example of toy models in which one can
 analyze the quantum nature of black holes using
 conventional conformal field theory.
 A matrix model for this black hole has also been
 proposed recently\cite{KKK}.
 It has also been used to analyze the theory of NS5-branes\cite{ABKS},
 or strings propagating in the vicinity of singularities of
 Calabi-Yau manifolds\cite{GKP,GK,GK2,ES,SY}
 in the spirit of holographic duality.
 The string theory on these backgrounds are supposed to be
 described by linear dilaton CFTs,
 but typically the dilaton diverges at the points where
 the NS5-branes or the singularity of Calabi-Yau manifolds
 are located.
 The use of the coset CFT therefore serves as a regularization.
 The coset model also has an application to the analysis
 of the $SL(2,{\bf R})$ WZW model,
 where the processes involving ``long strings'' have been studied
 in recent works\cite{MO,MOS,GN,GN2}.

   Fateev, Zamolodchikov and Zamolodchikov(FZZ) have made
 a conjecture that the above coset model has a dual description
 in terms of the sine-Liouville theory\cite{FZZ}.
 They checked the conjecture from various approaches,
 including the explicit comparison of two- and three-point
 functions in both theories.
 Based on this duality, there have been a lot of analysis
 of string theory on singular Calabi-Yau manifolds,
 using the supersymmetric version of sine-Liouville theory
 to describe the deformed singularities.
 It should also be noted that the matrix model\cite{KKK}
 has also been proposed based upon this conjecture.
 So it is of great importance to know this duality in detail.
 Recently, \cite{HK} has made an analysis
 using the linear sigma model techniques and concluded
 that the supersymmetric version of these two theories
 are dual to each other.
 To analyze the coset model from the dual sine-Liouville
 theory, it would also be very useful if we have explicit
 expressions for correlators in the latter theory.

   In this paper we study the three-point functions in
 sine-Liouville theory on a sphere.
 In a perturbative approach they can be calculated
 as free-field correlators with a number of insertions
 of ``screening operators''.
 So we can express them as certain multiple integrals
 over the complex plane.
 However, as we will see, to evaluate the integral expression
 we will need to develop some techniques and have a new
 formula like the one in \cite{DF}.

   This paper is organized as follows. 
 Section 2 is a brief introduction to
 sine-Liouville theory, in which we shall give an integral
 expression for generic three-point functions of tachyons.
 In section 3 we undertake a detailed analysis of
 the resultant integral expression.
 We utilize the techniques of \cite{DF,D} for translating
 $d^2z$ integrals into products of contour integrals,
 and evaluating the resultant contour integrals.
 Along the way we prove that the three-point correlator
 vanishes unless the total winding number is either $0$ or $\pm 1$.
 This is in consistency with the result of FZZ
 that in generic $N$-point functions the winding number
 conservation is violated up to $N-2$ units.
 We finally end up with some new formulae and closed analytic
 expressions for generic three-point functions of tachyons.
 Section 4 compares our expression for the correlators in
 sine-Liouville theory with the known formula in the coset model.
 In the case where the total winding number is zero, we see that
 the two expressions are in good agreement,
 and the $SL(2,{\bf R})$ group-theoretical factor is correctly
 reproduced in sine-Liouville theory.
 Using our new formulae, in section 5 we analyze the three-point
 amplitudes in superstring theory on a linear-dilaton background
 which is described by the $N=2$ super-Liouville theory.
 The last section concludes with a brief discussion.
 Our new formulae are summarized in the appendix for the readers' use.

\section{Sine-Liouville Theory}
   Consider the two-dimensional free CFT with the action
\begin{equation}
  S = \frac{1}{2\pi}\int d^2z(
   \partial x\bar{\partial}x + \partial\rho\bar{\partial}\rho
   -\frac{\rho}{2\sqrt{2k'}}\sqrt{g}R),
\end{equation}
 with $k'=k-2$.
 If we introduce the interaction of the form
\begin{equation}
  S_{\rm int} = \frac{\lambda}{2\pi}\int d^2z
  e^{-\sqrt{k'/2}\,\rho}\cos\sqrt{k/2}(x_L-x_R)
\end{equation}
 which is made of the screening currents
\begin{equation}
  \eta^\pm = \exp[\pm i\sqrt{k/2}\,x -\sqrt{k'/2}\,\rho],
\end{equation}
 the resultant theory is conjectured to give
 a dual description of the $SL(2,{\bf R})/U(1)$ coset model.
 In this conjecture, the primary fields are
 described by tachyonic vertex operators
\begin{equation}
  T_{jm\bar{m}}(z,\bar{z})
 =T_{jm}(z)T_{j\bar{m}}(\bar{z})
 = \exp[j\sqrt{2/k'}\rho +i\sqrt{2/k}\,(mx_L+\bar{m}x_R)],
\end{equation}
 satisfying the quantization condition
\begin{equation}
  m = \frac{1}{2}(p+kw),~~~
 \bar{m} = \frac{1}{2}(p-kw).
\end{equation}
 The integers $p$ and $w$ are the momentum and
 the winding number along the periodic direction
 of the Euclidean two-dimensional black hole(cigar).
 It is expected that in generic correlation functions
 in the coset model the winding number do not conserve,
 while the momentum does conserve.
 It is therefore interesting to study how this property
 is realized in the dual sine-Liouville theory.
 It involves the calculation of correlation functions
 in free CFT of the form:
\begin{equation}
  \frac{1}{m!n!}\left<\prod_a T_{j_am_a\bar{m}_a}(w_a)
         \left(\int d^2z S_{\rm int}^+(z)\right)^m
         \left(\int d^2z S_{\rm int}^-(z)\right)^n  \right>
\label{fcorr}
\end{equation}
 where the two screening operators $S_{\rm int^{\pm}}$ are defined by
\begin{equation}
  S_{\rm int}^+(z,\bar{z}) = \eta_L^+(z)\eta_R^-(\bar{z}),~~
  S_{\rm int}^-(z,\bar{z}) = \eta_L^-(z)\eta_R^+(\bar{z}).
\end{equation}

   We are particularly interested in
 three-point functions on the sphere.
 From the momentum conservation law of free CFT
 with background charge we have
\begin{equation}
  \sum_{a=1}^3 j_a +1    = \frac{k'}{2}(n+m),~~
  \sum_{a=1}^3 m_a       = \frac{k}{2}(n-m), ~~
  \sum_{a=1}^3 \bar{m}_a = \frac{k}{2}(m-n),
\end{equation}
 in accord with the momentum conservation
 and the winding number non-conservation,
\begin{equation}
  \sum p_a=0,~~ \sum w_a=n-m.
\end{equation}
 Using the free field OPEs we evaluate (\ref{fcorr}) and obtain
\begin{eqnarray}
&&I= \frac{1}{m!n!}\prod_{a<b}
         w_{ab}^{\frac{2m_am_b}{k} - \frac{2j_aj_b}{k'}}
         \bar{w}_{ab}^{\frac{2\bar{m}_a\bar{m}_b}{k} - \frac{2j_aj_b}{k'}}
 \nonumber \\ && \hskip1cm\cdot \prod_a
         \int d^{2m}z_id^{2n}z_{\hat{i}}
         \prod_{i<j}^m|z_{ij}|^{2}
         \prod_{\hat{i}<\hat{j}}^n|z_{\hat{i}\hat{j}}|^{2}
         \prod_{i,\hat{j}}^{m,n}|z_{i\hat{j}}|^{2-2k}
 \nonumber \\ && \hskip1cm\cdot \prod_a
   \left[  \prod_{i=1}^m
           (z_i-w_a)^{j_a+m_a}
           (\bar{z}_i-\bar{w}_a)^{j_a-\bar{m}_a}
           \prod_{\hat{i}=\hat{1}}^{\hat{n}}
           (z_{\hat{i}}-w_a)^{j_a-m_a}
           (\bar{z}_{\hat{i}}-\bar{w}_a)^{j_a+\bar{m}_a}
   \right].
\end{eqnarray}
 For three-point functions, the $w_a$-dependence
 is determined from the conformal invariance as
\begin{equation}
  I=\prod_{a<b}^3 w_{ab}^{-h_{ab}}\bar{w}_{ab}^{-\bar{h}_{ab}}
  D(j_a,m_a,\bar{m}_a)
\end{equation}
 where we introduced the notation such as $h_{12}= h_1+h_2-h_3$.
 $h_a$ denotes the conformal weight of $T_{j_am_a\bar{m}_a}$
 which is given by
\begin{equation}
  h       = -\frac{j(j+1)}{k'}+\frac{m^2}{k},~~
  \bar{h} = -\frac{j(j+1)}{k'}+\frac{\bar{m}^2}{k}.
\end{equation}
 Then all we need is the integral formula of the form:
\begin{eqnarray}
 I(\alpha,\beta,\gamma)
 &=& \frac{1}{m!n!}\int\prod_{i=1}^m\left(
      d^2z_i z_i^\alpha\bar{z}_i^{\bar{\alpha}}
           (z_i-1)^\beta(\bar{z}_i-1)^{\bar{\beta}}\right) 
     \prod_{i<j}^m|z_{ij}|^2
    \nonumber \\ && ~~~~~~\cdot
   \prod_{\hat{i}=\hat{1}}^n\left(
      d^2z_{\hat{i}}
      z_{\hat{i}}^{\alpha'}\bar{z}_{\hat{i}}^{\bar{\alpha}'}
      (z_{\hat{i}}-1)^{\beta'}(\bar{z}_{\hat{i}}-1)^{\bar{\beta}'}\right) 
     \prod_{{\hat{i}}<{\hat{j}}}^n|z_{\hat{i}\hat{j}}|^2
     \prod_{i,\hat{j}}^{m,n}|z_{i\hat{j}}|^{2-2k}.
\label{Iabc}
\end{eqnarray}
 This must be symmetric in $\alpha,\beta$ and $\gamma$
 obeying the relation
\begin{equation}
\begin{array}{ccccc}
     \alpha+\beta+\gamma
 &=& \bar{\alpha}+\bar{\beta}+\bar{\gamma}
 &=& kn-m-n-1 ,\\
     \alpha'+\beta'+\gamma'
 &=& \bar{\alpha}'+\bar{\beta}'+\bar{\gamma}'
 &=& km-m-n-1,
\end{array}
\end{equation}
 reflecting the symmetry under the exchange of three vertices.
 Once the above integral is calculated, the three-point functions
 can be obtained by the replacement
\begin{equation}
\begin{array}{llll}
 \alpha       =j_1+m_1, &
 \alpha'      =j_1-m_1, &
 \bar{\alpha} =j_1-\bar{m}_1, &
 \bar{\alpha}'=j_1+\bar{m}_1, \\
 \beta        =j_2+m_2, &
 \beta'       =j_2-m_2, &
 \bar{\beta}  =j_2-\bar{m}_2, &
 \bar{\beta}' =j_2+\bar{m}_2, \\
 \gamma       =j_3+m_3, &
 \gamma'      =j_3-m_3, &
 \bar{\gamma}=j_3-\bar{m}_3, &
 \bar{\gamma}'=j_3+\bar{m}_3.
\end{array}
\label{jmabg}
\end{equation}

\section{Integral Formula}
   Here we shall give the formula for the integral
\begin{eqnarray}
 I&=& \frac{1}{m!n!}\int\prod_{i=1}^m\left(
      d^2z_i z_i^\alpha\bar{z}_i^{\bar{\alpha}}
           (1-z_i)^\beta(1-\bar{z}_i)^{\bar{\beta}}\right) 
     \prod_{i<j}^m|z_{ij}|^{4\rho}
    \nonumber \\ && ~~~~~~\cdot
   \prod_{\hat{i}=\hat{1}}^n\left(
      d^2z_{\hat{i}}
      z_{\hat{i}}^{\alpha'}\bar{z}_{\hat{i}}^{\bar{\alpha}'}
      (1-z_{\hat{i}})^{\beta'}(1-\bar{z}_{\hat{i}})^{\bar{\beta}'}\right) 
     \prod_{{\hat{i}}<{\hat{j}}}^n|z_{\hat{i}\hat{j}}|^{4\rho'}
     \prod_{i,\hat{j}}^{m,n}|z_{i\hat{j}}|^{4\sigma},
\end{eqnarray}
 with $\rho=\rho'=1/2$.

\subsection{Decomposing into two contour integrals}
   The first thing to do is to transform it
 into a multiple contour integral.
 To do this, we rewrite the above expression
 in terms of real variables $(x_i,y_i,x_{\hat{i}},y_{\hat{i}})$
\begin{eqnarray}
  I &=& \frac{1}{m!n!}\int\prod_{i=1}^m\left[
      dx_idy_i (x_i+iy_i)^\alpha(x_i-iy_i)^{\bar{\alpha}}
               (1-x_i-iy_i)^\beta(1-x_i+iy_i)^{\bar{\beta}}\right]
    \nonumber \\ && ~~~~~~~~~
    \prod_{\hat{i}=\hat{1}}^n\left[
      dx_{\hat{i}}dy_{\hat{i}}
      (x_{\hat{i}}+iy_{\hat{i}})^{\alpha'}
      (x_{\hat{i}}-iy_{\hat{i}})^{\bar{\alpha}'}
      (1-x_{\hat{i}}-iy_{\hat{i}})^{\beta'}
      (1-x_{\hat{i}}+iy_{\hat{i}})^{\bar{\beta}'} \right]
    \nonumber \\ && ~~~~~~~~~
     \prod_{i<j}^m\left(x_{ij}^2+y_{ij}^2\right)^{2\rho}
     \prod_{\hat{i}<\hat{j}}^n
     \left(x_{\hat{i}\hat{j}}^2+y_{\hat{i}\hat{j}}^2\right)^{2\rho'}
    \prod_{i,\hat{j}}^{m,n}
    \left(x_{i\hat{j}}^2+y_{i\hat{j}}^2\right)^{2\sigma},
\end{eqnarray}
 Next we rotate the integration contours of $y_i$ and $y_{\hat{i}}$
 by $-(90-\epsilon)$ degrees.
 This can be done by introducing new variables
\[
 y_i         = -i(1+i\epsilon)\tilde{y}_i,~~
 y_{\hat{i}} = -i(1+i\epsilon)\tilde{y}_{\hat{i}}
\]
 and make the change of the integration variables.
 Then we get (omitting tildes)
\begin{eqnarray}
  I &=& \frac{1}{m!n!}\int
   \prod_{i=1}^m\left[
      -idx_idy_i
        (x_i+y_i+i\epsilon y_i)^\alpha
        (x_i-y_i-i\epsilon y_i)^{\bar{\alpha}}
    \right. \nonumber \\ && ~~~~~~~~~~~~~~~~~~~~~~~ \left.
        (1-x_i-y_i-i\epsilon y_i)^\beta
        (1-x_i+y_i+i\epsilon y_i)^{\bar{\beta}}\right]
    \nonumber \\ && ~~~~~~~
    \cdot\prod_{\hat{i}=1}^{\hat{n}}\left[
      -idx_{\hat{i}}dy_{\hat{i}}
        (x_{\hat{i}}+y_{\hat{i}}+i\epsilon y_{\hat{i}})^\alpha
        (x_{\hat{i}}-y_{\hat{i}}-i\epsilon y_{\hat{i}})^{\bar{\alpha}}
    \right. \nonumber \\ && ~~~~~~~~~~~~~~~~~~~~~~~ \left.
        (1-x_{\hat{i}}-y_{\hat{i}}-i\epsilon y_{\hat{i}})^\beta
        (1-x_{\hat{i}}+y_{\hat{i}}+i\epsilon y_{\hat{i}})^{\bar{\beta}}\right]
    \nonumber \\ && ~~~~~~~
    \cdot\prod_{i<j}^m\left[x_{ij}^2-(1+i\epsilon)y_{ij}^2\right]^{2\rho}
     \prod_{\hat{i}<\hat{j}}^{\hat{n}}
     \left[x_{\hat{i}\hat{j}}^2-(1+i\epsilon)
                       y_{\hat{i}\hat{j}}^2\right]^{2\rho'}
     \prod_{i,\hat{j}}^{m,\hat{n}}
     \left[x_{i\hat{j}}^2-(1+i\epsilon)y_{i\hat{j}}^2\right]^{2\sigma},
\label{rot}
\end{eqnarray}
 where all the variables take values on the real line.
 We further introduce new integration variables
\[
 z_i = x_i +y_i,~~
 w_i = x_i -y_i,~~
 z_{\hat{i}} = x_{\hat{i}} +y_{\hat{i}},~~
 w_{\hat{i}} = x_{\hat{i}} -y_{\hat{i}}.
\]
 to rewrite the integral as follows:
\begin{eqnarray}
 I &=& \frac{1}{m!n!}\cdot\nonumber \\
   & & \int
   \prod_{i=1}^m\left[
      \frac{idz_idw_i}{2}
          \left(z_i-i\epsilon w_i\right)^\alpha
          \left(w_i-i\epsilon z_i\right)^{\bar{\alpha}}
          \left(1-z_i-i\epsilon(1-w_i)\right)^\beta
          \left(1-w_i-i\epsilon(1-z_i)\right)^{\bar{\beta}} \right]
    \nonumber \\ && ~\cdot
   \prod_{\hat{i}=\hat{1}}^n\left[
      \frac{idz_{\hat{i}}dw_{\hat{i}}}{2}
          \left(z_{\hat{i}}-i\epsilon w_{\hat{i}}\right)^{\alpha'}
          \left(w_{\hat{i}}-i\epsilon z_{\hat{i}}\right)^{\bar{\alpha'}}
          \left(1-z_{\hat{i}}-i\epsilon(1-w_{\hat{i}})\right)^{\beta'}
          \left(1-w_{\hat{i}}-i\epsilon(1-z_{\hat{i}})\right)^{\bar{\beta'}}
       \right]
    \nonumber \\ && ~\cdot
     \prod_{i<j}^m\left[
          \left(z_{ij}-i\epsilon w_{ij}\right)^{2\rho}
          \left(w_{ij}-i\epsilon z_{ij}\right)^{2\rho} \right]
     \prod_{\hat{i}<\hat{j}}^n\left[
         \left(z_{\hat{i}\hat{j}}-i\epsilon w_{\hat{i}\hat{j}}\right)^{2\rho'}
         \left(w_{\hat{i}\hat{j}}-i\epsilon z_{\hat{i}\hat{j}}\right)^{2\rho'}
     \right]
    \nonumber \\ && ~\cdot
     \prod_{i,\hat{j}}^{m,n}\left[
          \left(z_{i\hat{j}}-i\epsilon w_{i\hat{j}}\right)^{2\sigma}
          \left(w_{i\hat{j}}-i\epsilon z_{i\hat{j}}\right)^{2\sigma} \right],
\end{eqnarray}
 where all the variables takes values on the real line.
 The integral is ``almost factorized''
 into two multiple contour integrals.
 The appearance of $\epsilon$ serves as the specification
 of how to arrange the contours.
 To see how the contours should be arranged, suppose that
 we are to calculate the above integral by first performing
 the integration over $w_i$ and $w_{\hat{i}}$
 with $(z_i,z_{\hat{i}})$ held fixed.
 So we first focus on the following part
\begin{eqnarray}
   & & \int d^{m+n}w
   \prod_{i=1}^m\left[
          \left(w_i-i\epsilon z_i\right)^{\bar{\alpha}}
          \left(1-w_i-i\epsilon(1-z_i)\right)^{\bar{\beta}} \right]
    \nonumber \\ && ~~~~~~~~~~\cdot
   \prod_{\hat{i}=\hat{1}}^n\left[
          \left(w_{\hat{i}}-i\epsilon z_{\hat{i}}\right)^{\bar{\alpha'}}
          \left(1-w_{\hat{i}}-i\epsilon(1-z_{\hat{i}})\right)^{\bar{\beta'}}
       \right]
    \nonumber \\ && ~~~~~~~~~~\cdot
     \prod_{i<j}^m\left[
          \left(w_{ij}-i\epsilon z_{ij}\right)^{2\rho} \right]
     \prod_{\hat{i}<\hat{j}}^n\left[
         \left(w_{\hat{i}\hat{j}}-i\epsilon z_{\hat{i}\hat{j}}\right)^{2\rho'}
     \right]
     \prod_{i,\hat{j}}^{m,n}\left[
          \left(w_{i\hat{j}}-i\epsilon z_{i\hat{j}}\right)^{2\sigma} \right].
\label{wpart}
\end{eqnarray}
 The factors in the first two lines determine how to
 deform the contours of $w_i$ and $w_{\hat{i}}$ to keep away
 from the singularities 0 and 1.
 For example, if $0 <z_1< 1$, then the contour of $w_1$
 should pass below 0 and above 1.
 The rule is summarized in the figure \ref{pass01}:
\fig{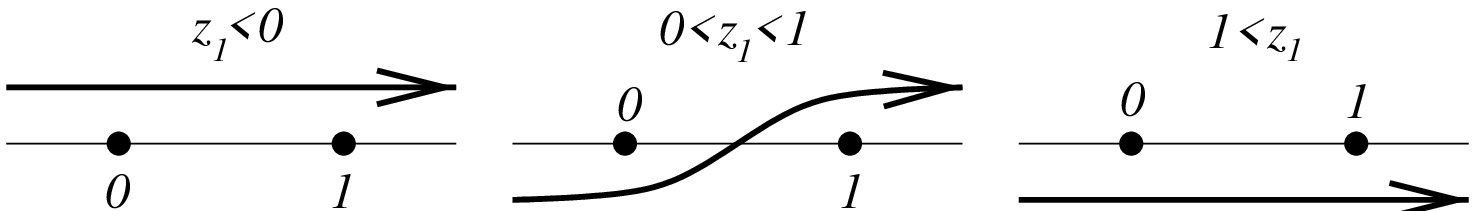}
    {The contour of $w_1$ depends on the value of $z_1$.}\\
 For more explanation of this manipulation of contours, see \cite{D}.
 In the same way, the third line determines how to align
 contours of $w_i,w_{\hat{i}}$ keeping away from one another.
 For example, if $z_1>z_2$, the contour of $w_1$ should be
 aligned below that of $w_2$.
 In this way the contours of $(w_i,w_{\hat{i}})$
 are aligned differently according to
 the values of $(z_i,z_{\hat{i}})$.
 Thus the integral $I$ is expressed as a sum
 over different orders of magnitude of $(z_i,z_{\hat{i}})$.
 Labeling different orders of $z$'s by $\tau$, it is expressed as
\begin{equation}
 I = \sum_\tau J(C_\tau,\alpha,\beta)\times
               J(P_\tau,\bar{\alpha},\bar{\beta}).
\label{tsum}
\end{equation}
 Here $C_\tau, P_\tau$ specify the
 domain of $z$-integration and contours of $w$'s.
 For example, for $\tau=(\hat{2}2\hat{1}1)$ we have to
 integrate over the domain
\begin{eqnarray*}
 0\le z_{\hat{2}} \le z_2 \le z_{\hat{1}} \le z_1 \le 1,
\end{eqnarray*}
 and the contours of $w$'s are arranged so that the contour of
 $w_{\hat{2}}$ lies on the top, and that of $w_1$ on the bottom,
 as in the figure \ref{order}.
 The reason why $z$'s are bounded between $0$ and $1$ is
 explained shortly.
\fig{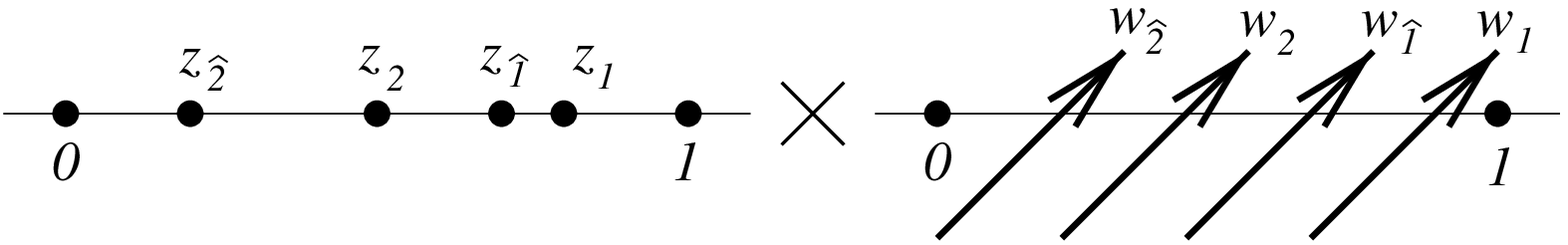}
    {Contribution from $\tau=(\hat{2}2\hat{1}1)$.}

   The integral $J(X_\tau,\alpha,\beta)$
 has the following form
\begin{equation}
 J(X_\tau,\alpha,\beta)=\int_{X_\tau}d^{m+n}z
  \prod_{i=1}^{m} z_i^\alpha(1-z_i)^\beta
  \prod_{\hat{i}=\hat{1}}^{\hat{n}}
    z_{\hat{i}}^{\alpha'}(1-z_{\hat{i}})^{\beta'}
  \left[\prod z_{ij}^{2\rho}
        \prod z_{\hat{i}\hat{j}}^{2\rho'}
        \prod z_{i\hat{j}}^{2\sigma}
  \right]_\tau.
\label{JX}
\end{equation}
 The last factor in the integrand is the $\tau$-dependent
 product of differences, which is positive when the order of
 $z$'s is precisely as dictated by $\tau$.
 An example is the following:
\begin{equation}
 \left[\prod z_{ij}^{2\rho}
       \prod z_{\hat{i}\hat{j}}^{2\rho'}
       \prod z_{i\hat{j}}^{2\sigma}
 \right]_{(\hat{2}2\hat{1}1)}
 = z_{12}^{2\rho}z_{\hat{1}\hat{2}}^{2\rho'}
   (z_{1\hat{1}}z_{1\hat{2}}z_{\hat{1}2}z_{2\hat{2}})^{2\sigma}.
\end{equation}
 For skeptical readers, we would like to note
 that the above definition of $J$ precisely accounts
 for the phases of the integrand of $I$
 in different regions in the complex $z,w$-space.
 Looking at (\ref{rot}) closely, we see that
 the integrand is positive when the order of $z$'s are
 the same as that of $w$'s
 (because we have, for instance,
  $z_{ij}w_{ij}=x_{ij}^2-y_{ij}^2>0$ in that case)
 and yields phases when there are relative flips.

   Once the arrangement of contours of $w$'s
 is determined according to the value of $z$'s,
 the contours can be deformed as long as
 they do not pass over the points $0,1$ or cross one another.
 Then we find that, if one of $z$'s is not in the range $[0,1]$,
 then at least one of the contours of $w$'s can slip off
 to infinity where the integral is assumed to vanish.
 So we may restrict the value of $z$'s to $[0,1]$ as was
 stated in the previous paragraph and
 implicitly assumed in the figure \ref{order}.

   The integral $J(P_\tau,\bar{\alpha},\bar{\beta})$ is given by
 the integrand of the form (\ref{JX}) and the contours of
 $w_i,w_{\hat{i}}$ arranged according to the order $\tau$.
 Remarkably, the integral vanishes for certain alignments of contours
 if we set $\rho=\rho'=1/2$.
 To see this, let us consider the following double integral
 as an example:
\begin{equation}
  \int_{-\infty}^{\infty} dw_1dw_2 (w_1-w_2)^{2\rho}f(w_1)f(w_2).
\end{equation}
 For generic $\rho$ this integral has a phase ambiguity
 due to the multi-valuedness of $(w_1-w_2)^{2\rho}$
 in the integrand.
 One has to remove this ambiguity by defining
 its value to be positive when $w_1>w_2$ and at the same time
 deforming the contours of $w_{1,2}$ off the real axis
 so that they do not cross.
 However, one does not encounter such problems in the case
 $\rho=1/2$ where the integral actually vanishes due to
 the permutation symmetry.
 Namely, the contribution from $w_1>w_2$ and that from $w_1<w_2$
 cancels each other.
 Returning to the $(w_i,w_{\hat{i}})$-integral
 $J(P_\tau,\bar{\alpha},\bar{\beta})$,
 we thus arrive at the following
\begin{quote}
 {\bf Lemma.}~~
 {\it If the multiple-contour $P_\tau$ contains
      two contours of $w_i$ and $w_j$
      (or those of $w_{\hat{i}}$ and $w_{\hat{j}}$)
      just next to each other, the integral vanishes.
      In other words, it is non-vanishing
      only when the contours of $w_i$ and those of $w_{\hat{i}}$
      are arranged alternately.}
\end{quote}
 This property dramatically reduces the sums over $\tau$
 to those of only one or two terms, up to permutations
 within $w_i$'s or within $w_{\hat{i}}$'s.
 There are two terms when $m=n$, only one when $m=n\pm 1$.
 Otherwise the integral itself vanishes.
 This states that the non-conservation of winding number
 is up to $\pm 1$ for three-point correlators,
 which is in accord with the result of FZZ.

   Let us summarize here the result up to now.
 We studied the integral
\begin{eqnarray}
 I&=& \frac{1}{m!n!}\int\prod_{i=1}^m\left(
      d^2z_i z_i^\alpha\bar{z}_i^{\bar{\alpha}}
           (1-z_i)^\beta(1-\bar{z}_i)^{\bar{\beta}}\right) 
     \prod_{i<j}^m|z_{ij}|^{4\rho}
    \nonumber \\ && ~~~~~~\cdot
   \prod_{\hat{i}=\hat{1}}^{n}\left(
      d^2z_{\hat{i}}
      z_{\hat{i}}^{\alpha'}\bar{z}_{\hat{i}}^{\bar{\alpha}'}
      (1-z_{\hat{i}})^{\beta'}(1-\bar{z}_{\hat{i}})^{\bar{\beta}'}\right) 
     \prod_{{\hat{i}}<{\hat{j}}}^n|z_{\hat{i}\hat{j}}|^{4\rho'}
     \prod_{i,\hat{j}}^{m,n}|z_{i\hat{j}}|^{4\sigma},
\end{eqnarray}
 with $\rho=\rho'=1/2$.
 This was shown to be non-vanishing only when $m-n=\pm 1$ or $0$.
 For $m=n+1$ we have
\begin{eqnarray}
 I &=& (i/2)^{m+n}\int_{C_\tau}d^{n+1}z_i d^nz_{\hat{i}}
       \prod_{i=1}^{n+1}
        z_i^\alpha(1-z_i)^\beta
       \prod_{\hat{i}={\hat{1}}}^{n}
        z_{\hat{i}}^{\alpha'}(1-z_{\hat{i}})^{\beta'}
       \left[
       \prod z_{ij}
       \prod z_{\hat{i}\hat{j}}
       \prod z_{i\hat{j}}^{2\sigma}
       \right]_\tau
  \nonumber \\ && \cdot
       \int_{P_\tau}d^{n+1}w_i d^nw_{\hat{i}}
       \prod_{i=1}^{n+1}
        w_i^{\bar{\alpha}}(1-w_i)^{\bar{\beta}}
       \prod_{\hat{i}={\hat{1}}}^{n}
        w_{\hat{i}}^{\bar{\alpha}'}(1-w_{\hat{i}})^{\bar{\beta}'}
       \left[
       \prod w_{ij}
       \prod w_{\hat{i}\hat{j}}
       \prod w_{i\hat{j}}^{2\sigma}
       \right]_\tau
\end{eqnarray}
 with $\tau=((n+1)\hat{n}n\cdots \hat{2}2\hat{1}1)$.
 For $m=n$ we have
\begin{eqnarray}
 I &=& (i/2)^{m+n}\int_{C_\tau}d^nz_i d^nz_{\hat{i}}
       \prod_{i=1}^n
        z_i^\alpha(1-z_i)^\beta
       \prod_{\hat{i}={\hat{1}}}^{n}
        z_{\hat{i}}^{\alpha'}(1-z_{\hat{i}})^{\beta'}
       \left[
       \prod z_{ij}
       \prod z_{\hat{i}\hat{j}}
       \prod z_{i\hat{j}}^{2\sigma}
       \right]_\tau
  \nonumber \\ && ~~~~\cdot
       \int_{P_\tau}d^n w_i d^nw_{\hat{i}}
       \prod_{i=1}^n
        w_i^{\bar{\alpha}}(1-w_i)^{\bar{\beta}}
       \prod_{\hat{i}={\hat{1}}}^{n}
        w_{\hat{i}}^{\bar{\alpha}'}(1-w_{\hat{i}})^{\bar{\beta}'}
       \left[
       \prod w_{ij}
       \prod w_{\hat{i}\hat{j}}
       \prod w_{i\hat{j}}^{2\sigma}
       \right]_\tau
  \nonumber \\ && \!\!\!\!\!\!
   +(\tau \rightarrow -\tau)
\end{eqnarray}
 with $\tau=(\hat{n}n\cdots \hat{2}2\hat{1}1)$
 and $-\tau$ the reversal of $\tau$.
 The combinatoric factor $(m!n!)^{-1}$ just cancels with
 the number of equivalent orders of integration variables.

\subsection{Evaluating the contour integral}
 We now turn to the evaluation of the integral
\begin{equation}
  J(X_\tau) = \int_X d^mz_i d^nz_{\hat{i}}
      \prod_{i=1}^m z_i^\alpha(1-z_i)^\beta
      \prod_{\hat{i}=\hat{1}}^{n}
      z_{\hat{i}}^{\alpha'}(1-z_{\hat{i}})^{\beta'}
      \left[
      \prod z_{ij}
      \prod z_{\hat{i}\hat{j}}
      \prod z_{i\hat{j}}^{2\sigma}
      \right]_\tau
\label{J_X}
\end{equation}
 with various contours $X_\tau$ and the order $\tau$.
 We define the integrand to be single-valued in the region
 $z_i,\hat{z}_{\hat{i}}~/\!\!\!\!\!\!\in[-\infty,0]\cup[1,\infty]$,
 and strictly positive when all $z$'s are on $(0,1)$
 with the order dictated by $\tau$. 
 For later convenience, we introduce the parameters
 $\gamma,\gamma'$ by the followiong relations
\begin{eqnarray*}
  \alpha+\beta+\gamma = -m-2\sigma n-1,~~~
  \alpha'+\beta'+\gamma' = -2\sigma m-n-1.
\end{eqnarray*}
 The parameters $\alpha,\alpha',\beta,\beta',\gamma,\gamma'$
 correspond to the quantum numbers of three tachyon vertices.
\fig{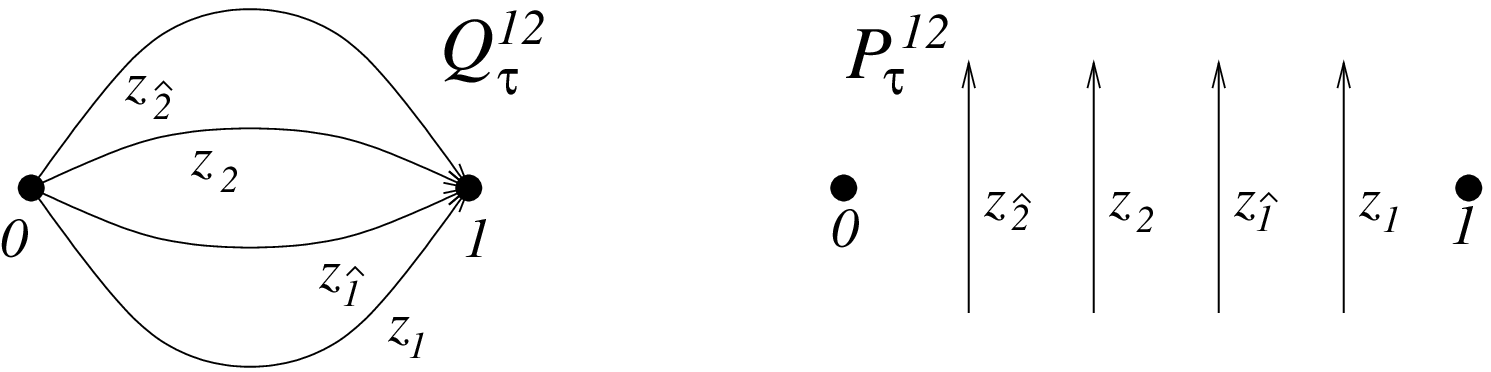}{Integrals $Q_\tau^{12}$ and $P_\tau^{12}$
           for $\tau=(\hat{2}2\hat{1}1)$.}

   To begin with, let us define some notation for integrals.
 Let us define $Q^{12}_\tau$ as the contour integral (\ref{J_X})
 with each contour originating from $0$ and heading for $1$.
 The contours are aligned according to $\tau$ and heading upwards.
 Similarly, define $P^{12}_\tau$ by the integral (\ref{J_X})
 with each contour passing between $0$ and $1$.
 Each contour sees $0$ on the left and $1$ on the right,
 and the contours are again aligned according to $\tau$.
 See the figure \ref{name} for a pictorial understanding.
 As a generalization, we define the integrals $P^{ab}_\tau$
 and $Q^{ab}_\tau$ by the interchange of the three vertices.
 Namely, they correspond to correlators with $a$-th and $b$-th
 vertices located at the points 0 and 1.
 Similarly, we define $C^{12}_\tau$ to be the integral
 $J(C_\tau)$ of the previous subsection, and define
 $C^{ab}_\tau$ by permutations of vertices.
 From these definition follows the relations
\begin{equation}
  P^{ab}_\tau = P^{ba}_{-\tau},~~
  Q^{ab}_\tau = Q^{ba}_{-\tau},~~
  C^{ab}_\tau = C^{ba}_{-\tau}.
\label{pqsym}
\end{equation}
 which hold without the phase ambiguity.
 One can also find that the contour $P_\tau$ of the previous
 section corresponds to $P_\tau^{12}$ in the new notation.

   Here we would like to note on the complex phases
 of the above integrals.
 Writing the contours of $P^{12}_\tau$ as in the right of the figure
 \ref{name} we find that the complex conjugation serves as
 the flips of the orientation of contours.
 Since this yields the factor $(-)^{m+n}$, it follows
 $P^{ab}_\tau$ is $i^{m+n}$ times a real quantity.
 $C^{ab}_\tau$ are real by definition. 

   Each of the above integrals is
 well-defined and analytic only in a finite region in
 the parameter space.
 For example, depending on the value of $\alpha$ and $\alpha'$,
 the integral $C^{12}_\tau$ may diverge due to the contribution
 from $z_i,z_{\hat{i}}\sim 0$.
 By a suitable approximate integration
 one finds when such divergence occurs.
 In the case $m=n$ we find that $C^{12}_\tau$ is
 well-defined when the real parts of
\[
 1+\alpha+\alpha'-nk',~
 1+\beta+\beta'-nk',~
 1+\alpha',~
 1+\beta
\]
 are positive, while in the case $m=n+1$
 the integral $C^{12}_\tau$ turns out to be well-defined when
\[
 1+\alpha+\alpha'-nk',~
 1+\beta+\beta'-nk',~
 1+\alpha,~
 1+\beta
\]
 have positive real parts\footnote{
 One might expect that the condition must be more strict
 in the case $m=n+1$, because more contours are ending on the points
 $0$ and $1$ as compared to the case $m=n$.
 Actually, using the techniques of manipulating contours
 which will be explained in the following,
 we may replace one of the $2n+1$ contours with the one
 which encloses all the other contours and has both ends
 on the same point ($0$ or $1$).
 So the singularities arise in the same way as in the case $m=n$.
 }.

 Using the integrals defined above, $I$ is expressed as
\begin{eqnarray}
  (m=n+1)~~~
  I &=& (i/2)^{m+n}C^{12}_\tau[\alpha,\beta]
                   P^{12}_\tau[\bar{\alpha},\bar{\beta}]
 , \nonumber \\
  (m=n)~~~
  I &=& (i/2)^{m+n}
   \left\{ C^{12}_\tau[\alpha,\beta]P^{12}_\tau[\bar{\alpha},\bar{\beta}]
          +C^{12}_{-\tau}[\alpha,\beta]P^{12}_{-\tau}[\bar{\alpha},\bar{\beta}]
   \right\}. 
\end{eqnarray}
 Thus the problem reduces to the calculation of contour integrals
 $P^{ab}_\tau,Q^{ab}_\tau$ and $C^{ab}_\tau$.
 To obtain them, we start with revealing the relations
 between $Q^{12}_\tau$ and $C^{12}_\tau$.
 Proceeding recursively we find,
\begin{eqnarray}
  Q^{12}_{(1)} &=&  C^{12}_{(1)}, 
  \nonumber \\
  Q^{12}_{(\hat{1}1)}&=&
  C^{12}_{(\hat{1}1)}+e^{-2\pi i\sigma}C^{12}_{(1\hat{1})},
  \\
  Q^{12}_{(2\hat{1}1)}
  &=& (1-e^{-4\pi i\sigma})C^{12}_{(2\hat{1}1)},
  \nonumber \\
  &\vdots \nonumber
\end{eqnarray}
 In this way we obtain the following relations:
\begin{eqnarray}
  (m=n+1)~~~
  Q^{12}_\tau &=& e^{i\pi n(n+1)k'/2}(-2i)^n
    \prod_{j=1}^n\sinpi{jk'}\; C^{12}_\tau,
    \nonumber \\
  (m=n)  ~~~
  Q^{12}_\tau &=& e^{i\pi n(n-1)k'/2}(-2i)^{n-1}
    \prod_{j=1}^{n-1}\sinpi{jk'}
    \left\{C^{12}_\tau -e^{i\pi nk'}C^{12}_{-\tau}\right\},
\end{eqnarray}
 where we used $k'=-2\sigma-1$ and $\sinpi{x}\equiv \sin(\pi x)$.

\fig{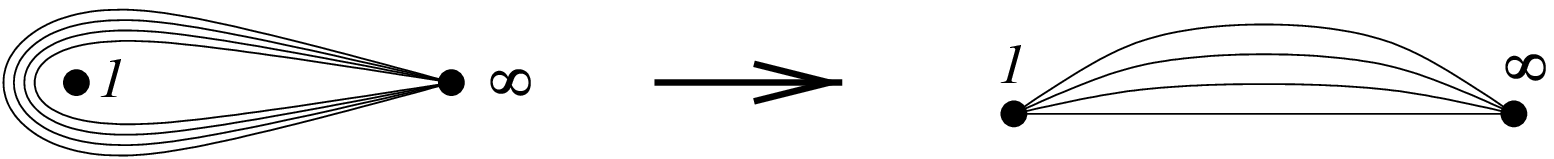}{Relation between $P^{12}_\tau$ and $Q^{23}_\tau$.}
   We can also find the relation between $P^{ab}_\tau$ and
 $Q^{ab}_\tau$ using a similar recursion.
 To do this, deform each of the contours in $P^{ab}_\tau$
 so that they all can be seen as starting from $\infty$,
 encircling around $1$ clockwise and ending on $\infty$
 as depicted in the left of the figure \ref{ptoq}.
 Dividing each contour into halves, we can relate
 the contour integral with another one, in which all the contours
 end on $1$ and $\infty$.
 After suitable change of variables we obtain the relation
 between $P^{12}_\tau$ and $Q^{23}_\tau$.
 For the first few cases it becomes
\begin{eqnarray}
  \tau=(1) \rightarrow ~~~
  P^{12}_\tau
   &=&  (e^{-i\pi\beta}-e^{i\pi\beta})Q^{23}_\tau
 \nonumber \\
   &=&  -2i\sinpi{\beta}C^{23}_\tau,
 \nonumber \\
  \tau=(\hat{1}1) \rightarrow ~~~
  P^{12}_\tau
   &=&  -2i\sinpi{\beta}
        \left\{e^{-i\pi\beta'}Q^{23}_\tau
              -e^{i\pi(\beta'+2\sigma)}Q^{23}_{-\tau}\right\}
 \nonumber \\
   &=& (-2i)^2\sinpi{\beta}\left\{
        \sinpi{\beta'}C^{23}_\tau + \sinpi{\beta'+2\sigma}C^{23}_{-\tau}
        \right\},
 \nonumber \\
  \tau=(2\hat{1}1) \rightarrow ~~~
  P^{12}_\tau
   &=& -2i\sinpi{\beta}(e^{-i\pi(\beta+\beta')}
                       -e^{i\pi(\beta+\beta'+4\sigma)})Q^{23}_\tau
 \nonumber \\
   &=& (-2i)^3\sinpi{\beta}\sinpi{k'}\sinpi{\beta+\beta'-k'}C^{23}_\tau,
 \nonumber \\
   &\vdots
\end{eqnarray}
 Proceeding in this way we obtain the relations
\begin{eqnarray}
\lefteqn{(m=n+1)} \nonumber \\
  P^{12}_\tau
 &=& (-2i)^{m}e^{-i\pi n(n+1)k'/2}\sinpi{\beta}
     \prod_{j=1}^n\sinpi{\beta+\beta'-jk'}Q^{23}_\tau
   \nonumber \\
 &=& (-2i)^{m+n}\sinpi{\beta}
     \prod_{j=1}^n\left[\sinpi{jk'}\sinpi{\beta+\beta'-jk'}\right]C^{23}_\tau,
   \nonumber \\
\lefteqn{(m=n)} \nonumber \\
   P^{12}_\tau
 &=& (-2i)^m e^{-i\pi n(n-1)k'/2}\sinpi{\beta}
     \prod_{j=1}^{n-1}\sinpi{\beta+\beta'-jk'}
  \left\{e^{-i\pi\beta'}Q^{23}_\tau+e^{i\pi(\beta'-nk')}Q^{23}_{-\tau}\right\}
   \nonumber \\
 &=& (-2i)^{m+n}\sinpi{\beta}
     \prod_{j=1}^{n-1}[\sinpi{jk'}\sinpi{\beta+\beta'-jk'}]
  \left\{\sinpi{\beta'}C^{23}_\tau
        -\sinpi{\beta'-nk'}C^{23}_{-\tau}\right\}.
\label{pcrel}
\end{eqnarray}
 These relations together with the symmetry (\ref{pqsym})
 largely determines the functional form of
 $P^{ab}_\tau, Q^{ab}_\tau$ and $C^{ab}_\tau$.
 We analyze the cases $m=n$ and $m=n+1$ separately in the following.

\subsection{The formula for $m=n+1$}
   First we study the simpler case $m=n+1$ where the only
 non-trivial $\tau$ is symmetric under the reversal, i.e.,
 $\tau=-\tau$.
 By exchanging the vertices in the previous relations we obtain
 the relations of the form:
\begin{equation}
  (i/2)^{m+n}P^{12}_\tau
 = C^{13}_\tau\,\sinpi{\alpha}\prod_{j=1}^n
   \sinpi{jk'}\sinpi{\alpha+\alpha'-jk'}
 = C^{23}_\tau\,\sinpi{\beta} \prod_{j=1}^n
   \sinpi{jk'}\sinpi{\beta  +\beta'-jk'} .
\end{equation}
 Using them together with the symmetry (\ref{pqsym})
 we can guess, as in \cite{DF},
\begin{eqnarray}
 C^{12}_\tau
 &=& \lambda_n(\sigma)
   \frac{\Gamma(1+\alpha)\Gamma(1+\beta)}{\Gamma(-\gamma)}
   \prod_{j=1}^n\left[\frac{\Gamma(1+\alpha+\alpha'-jk')
                            \Gamma(1+\beta+\beta'  -jk')
                          }{\Gamma(jk'-\gamma-\gamma')}\right] ,
 \nonumber \\
 P^{12}_\tau
 &=& \mu_n(\sigma)
   \frac{\Gamma(1+\gamma)}{\Gamma(-\alpha)\Gamma(-\beta)}
   \prod_{j=1}^n\left[\frac{\Gamma(1+\gamma+\gamma'-jk')
                          }{\Gamma(jk'-\alpha-\alpha')
                            \Gamma(jk'-\beta -\beta' )}\right],
 \nonumber \\
\lefteqn{
 (i/2)^{m+n}\mu_n(\sigma)
  = -\pi^{n+1}\lambda_n(\sigma)\prod_{j=1}^n\sinpi{-jk'}.
}\label{Jgues}
\end{eqnarray}
 The above form is also expected from the analytic structure
 of the integrals.
 In the previous subsection we discussed the region
 in the parameter space where the specific integrals
 are well-defined.
 In the same way we can study where in the parameter space
 the integrals have singularities.
 Taking the alignments of contours into account,
 we can convince ourselves that the singularities of the integrals
 in the parameter space are precisely as expressed
 in the above ansatz.
 It follows that $\lambda_n$ and $\mu_n$ have no poles in the
 parameter space, so they are independent of
 $\alpha,\alpha',\beta,\beta'$.
 Although our argument is not rigorous, we could in principle
 follow the same steps as in \cite{DF} to introduce many
 ``intermediate'' contour integrals that connect between
 $P^{ab}_\tau$'s and $C^{ab}_\tau$'s and
 do the analytic continuation strictly.
 In the following we would rather give some evidences
 supporting the above ansatz.

   We can find the asymptotic behavior of the expression
 at, say, large $\alpha,\alpha'$ with their ratio fixed.
 The integral $C^{12}_\tau$ behaves then like
\begin{eqnarray*}
  C^{12}_\tau
 &=& \int d^mz_i d^nz_{\hat{i}}
      \prod_{i=1}^m z_i^\alpha(1-z_i)^\beta
      \prod_{\hat{i}=\hat{1}}^{n}
      z_{\hat{i}}^{\alpha'}(1-z_{\hat{i}})^{\beta'}
      \left[
      \prod z_{ij}
      \prod z_{\hat{i}\hat{j}}
      \prod z_{i\hat{j}}^{2\sigma}
      \right]_\tau \\
 &=& \alpha^{-m-n}\int d^m\xi_i d^n\xi_{\hat{i}}\;
     e^{-(\Sigma_i\xi_i+\Sigma_{\hat{i}}\xi_{\hat{i}})/\alpha} 
      \prod_{i=1}^m e^{-\xi_i}(1-e^{-\xi_i/\alpha})^\beta
      \prod_{\hat{i}=\hat{1}}^{n}
      e^{-\alpha'\xi_{\hat{i}}/\alpha}(1-e^{-\xi_{\hat{i}}/\alpha})^{\beta'}
      \\ && ~~~~\cdot\left[
      \prod (e^{-\xi_i/\alpha}-e^{-\xi_j/\alpha})
      \prod (e^{-\xi_{\hat{i}}/\alpha}-e^{-\xi_{\hat{j}}/\alpha})
      \prod (e^{-\xi_i/\alpha}-e^{-\xi_{\hat{j}}/\alpha})^{2\sigma}
      \right]_\tau \\
 &\sim&
      \alpha^{-2n-1-n(\beta+\beta')-\beta-n^2-2n(n+1)\sigma}
      (1+{\cal O}(\alpha^{-1})).
\end{eqnarray*}
 On the other hand, using the asymptotic behavior of $\Gamma$
 function
\begin{equation}
  \Gamma(x)\sim \sqrt{2\pi}e^{-x}x^{x-1/2}
\end{equation}
 we find the asymptotic behavior of $C^{12}_\tau$
 guessed in (\ref{Jgues}) at large $\alpha,\alpha'$ is
\begin{eqnarray*}
  C^{12}_\tau
 &\sim&
 \alpha^{k'n(n+1)-\beta-1 -n(1+\beta+\beta')}
\end{eqnarray*}
 in precise agreement with the above.

   Another useful information on the integrals $P^{ab}_\tau$ 
 is the relation between $P^{ab}_\tau|_{\alpha\rightarrow -1}$
 and the one with fewer contours,
 from which we can determine the functional form of $\mu(\sigma)$.
 Let us consider $P^{12}_\tau$ in the limit $\alpha\rightarrow -1$,
 assuming that the leftmost variable in $\tau$ is without the hat.
 The limit can be evaluated by encircling the leftmost contour
 around $0$ and the rest around $1$, as in
 the figure \ref{sep}.
\fig{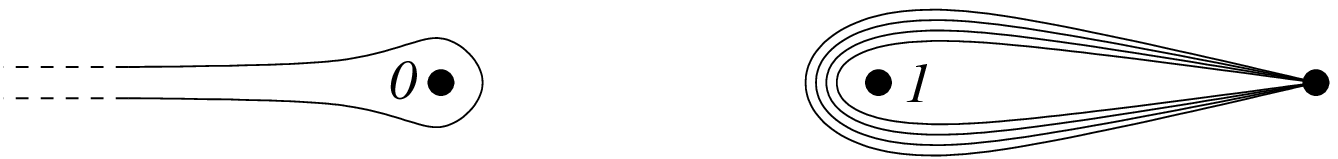}{The contours for evaluating the limit $\alpha\rightarrow -1$.}
 In the limit $\alpha\rightarrow -1$ the integral over the
 leftmost contour is well approximated by
 the contribution near $0$.
 At $\alpha = -1$ the leftmost contour can then be replaced by
 the one enclosing the single pole at $z=0$,
 and the integral is simply given by $2\pi i$ times the residue.
 Thus we have a relation between
 $P^{12}[m,n;\alpha,\alpha',\beta,\beta']$:
\begin{equation}
  P^{12}_\tau[m,n;-1,\alpha',\beta,\beta']
 =2\pi i P^{12}_\tau[m-1,n;0,\alpha'+2\sigma,\beta,\beta'].
\label{recur}
\end{equation}
 Applying this argument twice we find
\begin{equation}
  P^{12}_\tau[m,n;-1,-1-2\sigma,\beta,\beta']
 =-4\pi^2 P^{12}_\tau[m-1,n-1;2\sigma,0,\beta,\beta'].
\end{equation}
 This gives the recursion relation for $\mu_n(\sigma)$,
\begin{equation}
 \mu_n=\frac{-4\pi^2\mu_{n-1}(\sigma)}{\Gamma(1+nk')}.
\end{equation}
 Combined with $\lambda_0=1,\mu_0=2\pi i$ we obtain
\begin{equation}
  \mu_n = \frac{(2\pi i)^{2n+1}}{\prod_{j=1}^n\Gamma(1+jk')},~~
  \lambda_n = \prod_{j=1}^n\Gamma(-jk').
\end{equation}

   In all, the formula for $I$ in the case $m=n+1$ becomes:
\begin{eqnarray}
  I &=& -\pi^{m+n}(-)^{\gamma-\bar{\gamma}}
        \frac{\Gamma(1+\alpha)\Gamma(1+\beta)\Gamma(1+\gamma)}
             {\Gamma(-\bar{\alpha})\Gamma(-\bar{\beta})\Gamma(-\bar{\gamma})}
 \nonumber \\ && \cdot
        \prod_{j=1}^n
     \frac{\Gamma(-jk')
           \Gamma(1+\alpha+\alpha'-jk')
           \Gamma(1+\beta+\beta'-jk')
           \Gamma(1+\gamma+\gamma'-jk')}
          {\Gamma(1+jk')
           \Gamma(jk'-\alpha-\alpha')
           \Gamma(jk'-\beta-\beta')
           \Gamma(jk'-\gamma-\gamma')}.
\label{I_1}
\end{eqnarray}
 Rewriting $\alpha,\beta,\gamma$ in terms of
 the quantum numbers of coset CFT via (\ref{jmabg})
 and introducing the notation $\Delta(x)=\Gamma(x)/\Gamma(1-x)$,
 we obtain the following:
\begin{eqnarray}
 I  &=& -(-)^{m_3+\bar{m}_3}\pi^{m+n}
        \frac{\Gamma(m_1+j_1+1)\Gamma(m_2+j_2+1)\Gamma(m_3+j_3+1)}
             {\Gamma(\bar{m}_1-j_1)\Gamma(\bar{m}_2-j_2)\Gamma(\bar{m}_3-j_3)}
 \nonumber \\ && \cdot
        \left[\prod_{s=1}^n
             \Delta(1+sk')\Delta(sk'-2j_1)
             \Delta(sk'-2j_2)\Delta(sk'-2j_3)\right]^{-1}.
\end{eqnarray}
 We further rewrite this using the $\Upsilon$ function
 introduced in \cite{DO} and \cite{ZZ}.
 This is defined by
\begin{equation}
  \ln\Upsilon(x) = \int_0^\infty \frac{dt}{t}
  \left[ \left(\frac{Q}{2}-x\right)^2e^{-t}
         -\frac{\sinh^2[\left(\frac{Q}{2}-x\right)\frac{t}{2}]}
               {\sinh\frac{bt}{2}\sinh\frac{t}{2b}}
  \right],~~
  Q=b+b^{-1}.
\end{equation}
 We set the parameter $b$ as $b^{-2}=k'$.
 $\Upsilon(x)$ has zeroes at
\[
 x=-mb-nb^{-1},~ Q+mb+nb^{-1},~~  (m,n\in {\bf Z}_{\ge 0}).
\]
 Some useful properties are
\begin{eqnarray}
  \Upsilon(x+b) &=& \Delta(bx)b^{1-2bx}\Upsilon(x), \nonumber \\
  \Upsilon(x+b^{-1}) &=& \Delta(x/b)b^{2x/b-1}\Upsilon(x), \nonumber \\
  \Upsilon(Q-x) &=& \Upsilon(x),\nonumber \\
  \Upsilon'(-mb^{-1}) &=& (-)^mb^{-2m}\Upsilon[(m+1)b^{-1}]\Gamma(m+1)^2.
\end{eqnarray}
 Using this we can rewrite the above expression as
\begin{eqnarray}
  I &=& (-)^{n+1+m_3+\bar{m}_3}
        \frac{b^{2nk}\pi^{2n+1}
              \Gamma(m_1+j_1+1)\Gamma(m_2+j_2+1)\Gamma(m_3+j_3+1)}
             {\Gamma[b^2(\Sigma j_a +1)+\frac{1}{2}]^2
              \Gamma(\bar{m}_1-j_1)\Gamma(\bar{m}_2-j_2)\Gamma(\bar{m}_3-j_3)}
 \nonumber \\ && \cdot
        \frac{\Upsilon[b^{-1}]}{\Upsilon[(\Sigma j_a+1)b+\frac{1}{2b}]}
        \frac{\prod_{a=1}^3\Upsilon[(2j_a+1)b]}
             {\prod_{a<b}^3\Upsilon[-j_{ab}b+\frac{1}{2b}]}.
\end{eqnarray}
 where $n=b^2(\tsum j_a +1)-1/2$.

\subsection{The formula for $m=n$}
 The case $m=n$ is a little more complicated since
 there are essentially two non-equivalent alignments of contours,
 $\tau$ and $-\tau$.
 However, one can read off a lot from the relations (\ref{pcrel})
 between $P^{ab}_\tau$ and $C^{ab}_\tau$ as in the case $m=n+1$.
 The relations are simplified if we rescale
 $C^{ab}_\tau$ and $P^{ab}_\tau$ as
\begin{eqnarray}
 C^{12}_\tau
 &=& \lambda_n c^{12}_\tau\cdot
   \prod_{j=1}^{n-1}\left[
   \frac{\Gamma(1+\alpha+\alpha'-jk')\Gamma(1+\beta+\beta'  -jk')}
        {\Gamma(jk'-\gamma-\gamma')}\right] , \\
 (i/2)^{m+n}P^{12}_\tau
 &=& \pi^{n-1}\lambda_np^{12}_\tau\cdot
   \prod_{j=1}^{n-1}\left[
   \frac{\sinpi{-jk'}\Gamma(1+\gamma+\gamma'-jk')}
        {\Gamma(jk'-\alpha-\alpha')\Gamma(jk'-\beta-\beta')}\right].
\end{eqnarray}
 Using them we can rewrite (\ref{pcrel}) neatly in a matrix form:
\begin{equation}
 \IIxI{p^{12}_\tau}{p^{21}_\tau}
\!=\!A_\beta \IIxI{c^{23}_\tau}{c^{32}_\tau}
\!=\!A_\alpha^T\IIxI{c^{31}_\tau}{c^{13}_\tau},~
 \IIxI{p^{23}_\tau}{p^{32}_\tau}
\!=\!A_\gamma \IIxI{c^{31}_\tau}{c^{13}_\tau}
\!=\!A_\beta^T\IIxI{c^{12}_\tau}{c^{21}_\tau},~
 \IIxI{p^{31}_\tau}{p^{13}_\tau}
\!=\!A_\alpha \IIxI{c^{12}_\tau}{c^{21}_\tau}
\!=\!A_\gamma^T\IIxI{c^{23}_\tau}{c^{32}_\tau},
\label{c-rel}
\end{equation}
 where
\begin{equation}
  A_\alpha =\IIxII{\sinpi{\alpha}\sinpi{\alpha'}}
                  {-\sinpi{\alpha}\sinpi{\alpha'-nk'}}
                  {-\sinpi{\alpha'}\sinpi{\alpha-nk'}}
                  {\sinpi{\alpha}\sinpi{\alpha'}}
\end{equation}
 and $A_\beta,A_\gamma$ are defined similarly.

\paragraph{The case $m=n=1$}
   In proceeding further, let us analyze first the case $m=n=1$.
 As we will see, in this case we can perform
 the contour-integration rather explicitly.
 The result will be used to write down the formula
 for more generic cases.
 The integral $C^{12}_\tau$ is written in this case as
\begin{equation}
  C^{12}_\tau
 = \int_0^1 dz \int_0^z d\hat{z}
   z^{\alpha}(1-z)^\beta\hat{z}^{\alpha'}(1-\hat{z})^{\beta'}
   (z-\hat{z})^{2\sigma}.
\end{equation}
 Using the formulae involving (generalized) hypergeometric functions
\begin{eqnarray}
  \int_0^1 dt t^{b-1}(1-t)^{c-b-1}(1-zt)^{-a}
  &=& \frac{\Gamma(b)\Gamma(c-b)}{\Gamma(c)}
      F(a,b;c;z), \\
  \int_0^1 dt t^{c-1}(1-t)^{f-c-1}F(a,b;e;t)
  &=& \frac{\Gamma(c)\Gamma(f-c)}{\Gamma(f)}
     \ghyp(a,b,c;e,f;1),
\label{3F2-1}
\end{eqnarray}
 we can carry out the integration and obtain
\begin{eqnarray}
  C^{12}_\tau &=&
  \frac{\Gamma(\alpha+\alpha'-k'+1)\Gamma(\beta+1)
        \Gamma(\alpha'+1)\Gamma(-k')}
       {\Gamma(\alpha+\alpha'+\beta-k'+2)\Gamma(\alpha'-k'+1)}
  \nonumber \\ && \cdot
   \ghyp(-\beta',\alpha'+1,\alpha+\alpha'-k'+1;
          \alpha'-k'+1,\alpha+\alpha'+\beta-k'+2;1)
  \nonumber \\ &=&
  \frac{\Gamma(\alpha+\alpha'-k'+1)\Gamma(\beta+\beta'-k'+1)
        \Gamma(\alpha'+1)\Gamma(\beta+1)}
       {\Gamma(\alpha'-\gamma+1)\Gamma(\beta-\gamma'+1)}
  \nonumber \\ && \cdot
  \ghyp(\alpha'+1,\beta+1,k'-\gamma-\gamma';
        \alpha'-\gamma+1,\beta-\gamma'+1;1).
\end{eqnarray}
 In the above we repeatedly used the formula
\begin{equation}
  \ghyp(a,b,c;e,f;1) =
  \frac{\Gamma(e)\Gamma(f)\Gamma(s)}
       {\Gamma(a)\Gamma(s+b)\Gamma(s+c)}
  \ghyp(e-a,f-a,s;s+b,s+c;1)
\end{equation}
 with $s=e+f-a-b-c$, and the relations between parameters
 $\alpha+\beta+\gamma+1 = \alpha'+\beta'+\gamma'+1 = k'$.
 The next formula we use is the relation between
 $G\left[^{a,b,c}_{~e,f}\right]\equiv
  \frac{\Gamma(a)\Gamma(b)\Gamma(c)}{\Gamma(e)\Gamma(f)}
                 \ghyp(a,b,c;e,f;1)$:
\begin{eqnarray}
  G\IIxI{a,b,c}{e,f}
 &=& \frac{\sinpi{e-b}\sinpi{f-b}}{\sinpi{a}\sinpi{c-b}}
      G\IIxI{b,1+b-e,1+b-f}{1+b-c,1+b-a}
 \nonumber \\ &+&
 \frac{\sinpi{e-c}\sinpi{f-c}}{\sinpi{a}\sinpi{b-c}}
     G\IIxI{c,1+c-e,1+c-f}{1+c-b,1+c-a},
\end{eqnarray}
 from which we can derive the relations between $C^{ab}_\tau$
 of the form:
\begin{equation}
  \sinpi{\alpha'-\beta}C^{12}_\tau =
  \sinpi{\gamma}C^{13}_\tau -\sinpi{\gamma'}C^{32}_\tau.
\end{equation}
 Looking closely, the set of relations of this form turn to be
 equivalent to (\ref{pcrel}) or (\ref{c-rel}).
 Summarizing, in the case $m=n=1$ the integral $I$ is given by
\begin{equation}
 I = \int d^2z d^2w
     z^\alpha(1-z)^\beta
     \bar{z}^{\bar{\alpha}}(1-\bar{z})^{\bar{\beta}}
     w^{\alpha'}(1-w)^{\beta'}
     \bar{w}^{\bar{\alpha}'}(1-\bar{w})^{\bar{\beta}'}
     |z-w|^{4\sigma}.
\end{equation}
 It is be decomposed into contour integrals
\begin{equation}
  I = (i/2)^2\left\{
   C^{12}_\tau   [\alpha,\beta]P^{12}_\tau   [\bar{\alpha},\bar{\beta}]
  +C^{12}_{-\tau}[\alpha,\beta]P^{12}_{-\tau}[\bar{\alpha},\bar{\beta}]
             \right\}.
\end{equation}
 The integrals $C^{ab}_\tau$ are given by the formula
\begin{eqnarray}
 C^{12}_\tau &=&
  \frac{\Gamma(1+\alpha+\alpha'-k')\Gamma(1+\beta+\beta'-k')}
       {\Gamma(k'-\gamma-\gamma')}
 G\IIxI{\alpha'+1,\beta+1,k'-\gamma-\gamma'}{\alpha'-\gamma+1,\beta-\gamma'+1},
\label{c-tau}
\end{eqnarray}
 and $P^{ab}_\tau$ are related to $C^{ab}_\tau$ via
\begin{equation}
  (i/2)^2\IIxI{P^{12}_\tau}{P^{21}_\tau}
 = A_\beta\IIxI{C^{23}_\tau}{C^{32}_\tau}
 = A_\alpha^T\IIxI{C^{31}_\tau}{C^{13}_\tau},~~
 A_\alpha =
  \IIxII{\sinpi{\alpha}\sinpi{\alpha'}}{-\sinpi{\alpha}\sinpi{\alpha'-k'}}
        {-\sinpi{\alpha'}\sinpi{\alpha-k'}}{\sinpi{\alpha}\sinpi{\alpha'}}.
\end{equation}

\paragraph{Generic cases with $m=n$}

   Although it is extremely difficult to carry out
 the integration explicitly in generic cases,
 we find that the relations (\ref{c-rel}) among
 $c^{ab}_\tau$ and $p^{ab}_\tau$ have the same form
 as in the case $m=n=1$.
 If we assume $c^{12}_\tau$ to be given by replacing
 $k'$ with $nk'$ in (\ref{c-tau}), we have
\begin{equation}
 C^{12}_\tau
 = \lambda_n
  \prod_{j=1}^{n}\left[
   \frac{\Gamma(1+\alpha+\alpha'-jk')\Gamma(1+\beta+\beta'-jk')}
        {\Gamma(jk'-\gamma-\gamma')}  \right]
  G\IIxI{\alpha'+1,\beta+1, nk'-\gamma-\gamma'}
        {\alpha'-\gamma+1,\beta-\gamma'+1}
\end{equation}
 with $\lambda_1=1$.
 From the recursion relation (\ref{recur}) we obtain $\lambda_n$:
\begin{equation}
 \lambda_n = \prod_{j=1}^{n-1}\Gamma(-jk').
\end{equation}
 Putting all these together, we arrive at the following
 expression for $I$
\begin{eqnarray}
 I&=& \pi^{2n-2}
      \prod_{j=1}^{n-1}\left[
      \frac{\Gamma(-jk')
            \Gamma(1+\alpha+\alpha'-jk')
            \Gamma(1+\beta +\beta' -jk')
            \Gamma(1+\gamma+\gamma'-jk')}
           {\Gamma(1+jk')
            \Gamma(jk'-\alpha-\alpha')
            \Gamma(jk'-\beta -\beta' )
            \Gamma(jk'-\gamma-\gamma')}
     \right]
     \nonumber \\ &&
     \cdot \int d^2z d^2w
     z^\alpha (1-z)^{\beta}
     \bar{z}^{\bar{\alpha}} (1-\bar{z})^{\bar{\beta}}
     w^{\alpha'} (1-w)^{\beta'}
     \bar{w}^{\bar{\alpha}'} (1-\bar{w})^{\bar{\beta}'}
     |z-w|^{-2-2nk'}.
\end{eqnarray}
 Rewriting it further using $\Upsilon$ function and 
 $b={k'}^{-1/2}$ we obtain
\begin{equation}
  I = I_1(j_a,m_a,\bar{m}_a)
  \frac{(i\pi b^2)^{2n-2}}{\Gamma[b^2(\Sigma j_a +1)]^2}
  \frac{\Upsilon[b]\prod_{a=1}^3\Upsilon[(2j_a+1)b]}
       {\Upsilon[(\Sigma j_a +1)b]\prod_{a<b}^3\Upsilon[(j_{ab}+1)b]},
\label{I2}
\end{equation}
\begin{eqnarray}
  I_1(j_a,m_a,\bar{m}_a)
 &=& \int d^2z d^2w
   z^{j_1+m_1}(1-z)^{j_2+m_2}
   \bar{z}^{j_1-\bar{m}_1}(1-\bar{z})^{j_2-\bar{m}_2} 
 \nonumber \\ && ~~~~~~\cdot
   w^{j_1-m_1}(1-w)^{j_2-m_2}
   \bar{w}^{j_1+\bar{m}_1}(1-\bar{w})^{j_2+\bar{m}_2}
   |z-w|^{-2-2nk'}.
\label{Ijmm}
\end{eqnarray}

\section{A Check of the Duality}

   As was mentioned in the introduction, the sine-Liouville theory
 is believed to be dual to the $SL(2,{\bf R})/U(1)$ coset model.
 We check this duality by comparing the three-point functions
 of primary operators in the two theories.

   The three-point function of primary fields in the coset model
 is obtained from the correlation functions of primary fields
 $\Phi_j(z,\bar{z};x,\bar{x})$ in $SL(2,{\bf C})/SU(2)$ WZW model.
 $\Phi_j$ is characterized by the property
\begin{equation}
  \Phi_{jm\bar{m}}(z,\bar{z}) = \int d^2x x^{j+m}\bar{x}^{j+\bar{m}}
  \Phi_j(z,\bar{z};x,\bar{x})
\end{equation}
 where $\Phi_{jm\bar{m}}$ are primary fields
 of $SL(2,{\bf R})$-spin $j$ and $(J^3,\bar{J}^3)=(m,\bar{m})$.
 As shown in \cite{T} and also in \cite{HOS}, the three-point
 function of $\Phi_j$ is given by
\begin{eqnarray}
\lefteqn{
 \vev{\prod_{a=1}^3\Phi_{j_a}(z_a;x_a)}
 = D(j_a)\prod_{a<b}^3|z_{ab}|^{-2h_{ab}}|x_{ab}|^{-2j_{ab}-2}, }\nonumber \\
 D(j_a) &=& \frac{b^2\pi}{2}
  \frac{[k^{-1}b^{-2b^2}\Delta(b^2)]^{-\Sigma j_a -2}
        \Upsilon[b]\Upsilon[(2j_1+2)b]\Upsilon[(2j_2+2)b]\Upsilon[(2j_3+2)b]}
       {\Upsilon[(\Sigma j_a+2)b]
        \Upsilon[(j_{12}+1)b]\Upsilon[(j_{23}+1)b]\Upsilon[(j_{13}+1)b]}.
\end{eqnarray}
 Comparing this with (\ref{I2}), we find that the part that
 depends only on $j_a$ is in a good agreement\footnote{
  Actually the agreement is almost precise if we take into account
  the effect of wave-function renormalization of $\Phi_j$.
  Note also that the three-point function in sine-Liouville theory
  has an additional factor $\sim\Gamma(-2n)\Gamma(1+2n)$
  which arises from the integration over the $\rho$-zeromode
  and the expansion of $(S_{\rm int}^++S_{\rm int}^-)^{2n}$.
}.
 As for the part that depends also on $m_a$ and $\bar{m}_a$,
 the three-point function in coset model has the following factor
\begin{equation}
  \int \prod_{a=1}^3 [d^2x_a x_a^{j_a+m_a}\bar{x}_a^{j_a-\bar{m}_a}]
       \prod_{a<b}^3 |x_{ab}|^{-2j_{ab}-2}
 \sim \delta(\tsum m_a)\delta(\tsum \bar{m}_a) F(j_a,m_a,\bar{m}_a),
\end{equation}
 where the function $F$ is the left-right combined
 version of the $3j$-symbol: 
\begin{eqnarray}
 F(j_a,m_a,\bar{m}_a) &\equiv&
 \int d^2x_1 d^2x_2
   x_1^{j_1+m_1}\bar{x}_1^{j_1-\bar{m}_1}
   x_2^{j_2+m_2}\bar{x}_2^{j_2-\bar{m}_2}
 \nonumber \\ && ~~~~~~~~~~~~\cdot
   |1-x_1|^{-2j_{13}-2}|1-x_2|^{-2j_{23}-2}|x_{12}|^{-2j_{12}-2}.
\end{eqnarray}
 which can also be written in terms of generalized
 hypergeometric functions.
 The explicit form is found in \cite{HS}.
 So it represents the group-theoretical structure
 of the correlation functions.
 Quite surprisingly, using the formulae involving generalized
 hypergeometric functions we can show that $F$ precisely
 coincides with $I_1$ given in (\ref{Ijmm}).
 Hence we conclude that, in the case where the winding number
 conserves, the three-point functions in sine-Liouville theory
 agree with those in the coset model, including the $SL(2)$
 group-theoretical coefficients.

\section{An Application}

   Here we would like to apply our new formula to a problem
 in superstring theory on a singular Calabi-Yau(CY) manifold.
 Suppose we compactify the superstring theory on a CY manifold
 $X$ with a singularity.
 According to the holographic duality of \cite{ABKS,GKP},
 in the decoupling limit $g_s\rightarrow 0$ we end up with
 a non-gravitational but non-trivial interacting
 theory on the remaining non-compact space.
 As a typical case, if we take as $X$ an open non-compact
 CY two-fold with $A_k$ type singularity, the resultant
 theory in the decoupling limit is believed to describe
 the non-trivial field theory on $k$ coincident NS5-branes\cite{OV}.
 Generically the holographic duality states
 the correspondence between
\begin{eqnarray*}
\lefteqn{
 \mbox{decoupling limit of superstring on }{\bf R}^d\times X_{n}(F=0)}
 \nonumber \\
\lefteqn{
 (=\mbox{a non-trivial field theory on }{\bf R}^d)}
 \nonumber \\
 &\Longleftrightarrow&
 \mbox{superstring on }{\bf R}^d\times 
 {\bf R}_\rho \times S^1_Y \times LG(W=F).
\end{eqnarray*}
 Here $X_n$ is a CY $n$-fold ($2n+d=10$) defined by
 the complex equation $F(X)=0$, and $LG$ denotes the corresponding
 Landau-Ginzburg model with the superpotential $W=F$.
 Using this correspondence we can study the non-trivial field
 theories from the dual pictures.
 Here we shall calculate the three-point amplitudes of certain
 tachyonic operators using our new formula.
 The two-point functions can also be calculated in a similar
 way\cite{FK}.

   The worldsheet theory of a superstring on
 ${\bf R}^d\!\times\! {\bf R}_\rho\!\times\!
   S^1_Y\!\times\! LG$
 is given by the action
\begin{equation}
  S=S_{{\bf R}^d}+S_{L}+S_{LG},
\end{equation}
 where $S_{{\bf R}^d}, S_{LG}$ are the actions
 for ${\bf R}^d$ and Landau-Ginzburg degrees of freedom,
 respectively.
 The remaining degrees of freedom corresponding to
 the coordinates $\rho,Y$ are described by the $N=2$ Liouville action 
\begin{eqnarray}
  S_L &=&
 \frac{1}{2\pi}\int d^2z
 (\partial\rho\bar{\partial}\rho +\partial Y \bar{\partial}Y
  -\frac{Q}{4}\rho\sqrt{g}R
  \nonumber \\ && ~~~~~~~~~~~~
  +\psi_\rho\bar{\partial}\psi_\rho+\psi_Y\bar{\partial}\psi_Y
  +\bar{\psi}_\rho\partial\bar{\psi}_\rho
  +\bar{\psi}_Y\partial\bar{\psi}_Y)
  + S_{\rm int}^+ + S_{\rm int}^-, \nonumber \\
 S_{\rm int}^\pm &=&
    \frac{\mu_\pm}{2\pi} 
    \int d^2z (\psi_\rho\pm i\psi_Y)(\bar{\psi}_\rho\pm i\bar{\psi}_Y)
    e^{-\frac{1}{Q}(\rho \pm iY)},
\end{eqnarray}
 where $Q\equiv\sqrt{2/k}$.
 The theory reduces to the free theory if
 the cosmological constant $\mu_\pm$ vanishes.
 In this case the $N=2$ superconformal algebra is generated by
 the currents
\begin{eqnarray}
  T &=& -\frac{1}{2}
  ( \partial\rho\partial\rho +Q\,\partial^2\rho
   +\partial Y\partial Y
   +\psi_\rho\partial\psi_\rho + \psi_Y\partial\psi_Y), \nonumber \\
  T_F^\pm &=&
  \frac{i}{2}  (\psi_\rho\pm i\psi_Y)\partial(\rho\mp iY)
  +\frac{iQ}{2}\partial(\psi_\rho \pm i\psi_Y), \nonumber \\
  J &=& -i\psi_\rho\psi_Y +iQ\partial Y.
\end{eqnarray}
 Note that the interaction $S_{\rm int}^\pm$ is made from
 the screening currents which have no singular OPEs with
 $N=2$ superconformal currents up to total derivatives.

   In the following we calculate the correlators of ``tachyons'',
 which are expressed in the $(-1,-1)$-picture as
\begin{equation}
  T(z,\bar{z})=
  \exp\left[-\varphi+ik_\mu X^\mu +Q(j\rho-imY_L +i\bar{m}Y_R)\right],
\label{tac1}
\end{equation}
 where $\varphi$ constitutes the system of bosonized
 super-ghost together with $\xi$ and $\eta$.
 We may regard $j$ and $m$ as the quantum numbers of
 the corresponding operators in the $SL(2,{\bf R})/U(1)$
 Kazama-Suzuki model\cite{KS}.
 The minus sign in front of $m$ is due to the fact
 that the Liouville interaction terms explicitly break
 the conservation of $Y$-momentum,
 which we want to relate to the winding number along the cigar.
 So the momentum $p$ and the winding number $w$
 along the $S^1$ of the cigar are given by
\begin{equation}
  m=\frac{1}{2}(p+kw),~~ \bar{m}= \frac{1}{2}(p-kw).
\end{equation}
 The on-shell condition requires the parameters to satisfy
\begin{equation}
  \frac{k_\mu k^\mu}{2}+\frac{m^2-j(j+1)}{k}=
  \frac{k_\mu k^\mu}{2}+\frac{\bar{m}^2-j(j+1)}{k}=
  \frac{1}{2}.
\end{equation}
 The GSO condition requires the physical vertices to
 have odd worldsheet fermion number.
 Since it is defined by the $U(1)$ charge of the $N=2$
 superconformal symmetry we have,
\begin{equation}
 m,\bar{m}\in \frac{k}{2}(2{\bf Z}+1).
\end{equation}
 The corresponding vertices in the $(0,0)$-picture
 become
\begin{equation}
  \tilde{T}(z,\bar{z})= [ik_\mu\psi_X^\mu+Qj\psi_\rho-iQm\psi_Y]
     [ik_\mu\bar{\psi}_X^\mu+Qj\bar{\psi}_\rho+iQ\bar{m}\bar{\psi}_Y]
     e^{ik_\mu X^\mu+Q(j\rho-im_iY_L+i\bar{m}_iY_R)}.
\end{equation}

   The three-point function of tachyons is defined by
\begin{eqnarray}
  A &\equiv&
  \vev{c\bar{c}T_1(z_1)\,c\bar{c}T_2(z_2)\,c\bar{c}\tilde{T}_3(z_3)}
  \nonumber \\
    &=& \int [D\rho DY D\psi_\rho D\psi_Y\cdots]\;J_3
    \prod_{i=1}^{3}c\bar{c}e^{ik_iX+Q(j_i\rho-im_iY_L+i\bar{m}_iY_R)}(z_i)
    \; e^{-S} ,
 \\
 J_3 &=& e^{-\varphi(z_1,\bar{z}_1)-\varphi(z_2,\bar{z}_2)}
 \nonumber \\ 
 &&\times 
 [ik_3\cdot\psi_X+Qj_3\psi_\rho-iQm_3\psi_Y](z_3)
 [ik_3\cdot\bar{\psi}_X+Qj_3\bar{\psi}_\rho+iQ\bar{m}_3\bar{\psi}_Y]
 (\bar{z}_3).
\end{eqnarray}
 Since the tachyons do not contain the fields in the LG
 part, we can integrate over them and simply ignore
 the LG part.
 As was discussed in \cite{GL} (see also \cite{DFK}),
 by integrating over the zero-mode of $\rho$ the correlator reduces to
 that of a free theory with a number of insertions of
 $S_{\rm int}\equiv S_{\rm int}^+ + S_{\rm int}^-$:
\begin{eqnarray}
  A &\sim&   Q \Gamma(-s) \int [D\rho DYD\psi_\rho D\psi_Y\cdots]\; J_3 
  \prod_{i=1}^{3}c\bar{c}e^{ik_iX+Q(j_i\rho-im_iY_L+i\bar{m}_iY_R)}(z_i)
  \nonumber \\ && ~~~~~\times
 (S_{\rm int})^s
  \exp\left[-S_{{\bf R}^{d}}-S_{L\;(\mu_\pm=0)}\right], \\
  s &=& \frac{2}{k}(\Sigma j_i +1). \nonumber
\end{eqnarray}
 From the expression of $S_{\rm int}$ we find that
 only a part of $J_3$ containing $\psi_\rho, \psi_Y$
 yields non-zero contribution to $A$.
 So if we bosonize $\psi_\rho,\psi_Y$ as
\begin{equation}
  \psi_\rho \pm i\psi_Y = \sqrt{2}e^{\pm iH_L},~~
  \bar{\psi}_\rho \pm i\bar{\psi}_Y = \sqrt{2}e^{\pm iH_R},
\end{equation}
 we may replace the above $J_3$ with 
\[
  J_3 = \frac{1}{k} e^{-\varphi(z_1,\bar{z_1})-\varphi(z_2,\bar{z_2})}
    \left[ (j_3-m_3)(j_3+\bar{m}_3)e^{iH}
         + (j_3+m_3)(j_3-\bar{m}_3)e^{-iH} \right](z_3,\bar{z}_3).
\]
 The three-point function is then written as a sum of two terms,
 $A\sim A_+ + A_-$, with
\begin{eqnarray}
 A_\pm &=&
  \delta^n(\tsum_i k_i)\delta(\tsum_i p_i)\delta(\tsum_i w_i \mp 1)
  \frac{(j_3\mp m_3)(j_3\pm\bar{m}_3)}{k}
  \frac{\Gamma(-s)s!}{(\frac{s+1}{2})!(\frac{s-1}{2})!}
   \nonumber \\ &&\times
 \vev{e^{-\varphi(z_1)-\varphi(z_2)\pm iH(z_3)}
      \prod_{i=1}^3 c\bar{c}e^{ik_iX+Q(j_i\rho-im_iY_L+i\bar{m}_iY_R)}(z_i)
      (S_{\rm int}^+)^{\frac{s\mp 1}{2}}(S_{\rm int}^-)^{\frac{s\pm1}{2}}
      }_{\rm free}, \\
 S_{\rm int}^\pm
 &=& \frac{\mu_\pm}{\pi}\int d^2z e^{\pm iH-\frac{1}{Q}(\rho\pm iY)}.
\end{eqnarray}
 It is non-zero only when the total winding number is $\pm 1$. 
 So, in this case, the restriction on the total winding number
 simply comes from the argument of the worldsheet fermion number.
 This fact was also pointed out in \cite{GK2}.

   Let us consider the case $\tsum w_i =1$.
 The three-point function becomes
\begin{eqnarray}
  A_+ &=& \delta^n(\tsum_i k_i)\delta(\tsum p_i)\delta(\tsum w_i-1)
  \frac{\pi(j_3-m_3)(j_3+\bar{m}_3)}{k\sinpi{2n+1}n!(n+1)!}
  \left(\frac{\mu_+}{\pi}\right)^{n}
  \left(\frac{\mu_-}{\pi}\right)^{n+1}
 \nonumber \\ &&\cdot
  |z_{13}z_{23}|^2 \prod_{i<j}^3
  z_{ij}^{k_i\cdot k_j-\frac{2}{k}(m_im_j-j_ij_j)}
  \bar{z}_{ij}^{k_i\cdot k_j-\frac{2}{k}(\bar{m}_i\bar{m}_j-j_ij_j)}
\nonumber \\&&\cdot
 \int d^{2n+2}y_i d^{2n}y_{\hat{i}}
 \prod_{i=1}^{n+1}
 \left[
 (y_i-z_1)^{j_1-m_1}
 (y_i-z_2)^{j_2-m_2}
 (y_i-z_3)^{j_3-m_3-1} \right.
\nonumber \\&&~~~~~~~~~~~~~~~~~~~~~~~~~\left.
 (\bar{y}_i-\bar{z}_1)^{j_1+\bar{m}_1}
 (\bar{y}_i-\bar{z}_2)^{j_2+\bar{m}_2}
 (\bar{y}_i-\bar{z}_3)^{j_3+\bar{m}_3-1}
 \right]
\nonumber \\&&~~~~~~~~~~~~~~~~~~~\cdot
 \prod_{\hat{i}=1}^{n}
 \left[
 (y_{\hat{i}}-z_1)^{j_1+m_1}
 (y_{\hat{i}}-z_2)^{j_2+m_2}
 (y_{\hat{i}}-z_3)^{j_3+m_3+1} \right.
\nonumber \\&&~~~~~~~~~~~~~~~~~~~~~~~~~~\left.
 (\bar{y}_{\hat{i}}-\bar{z}_1)^{j_1-\bar{m}_1}
 (\bar{y}_{\hat{i}}-\bar{z}_2)^{j_2-\bar{m}_2}
 (\bar{y}_{\hat{i}}-\bar{z}_3)^{j_3-\bar{m}_3+1}
 \right]
\nonumber \\&&~~~~~~~~~~~~~~~~~~~\cdot
 \prod_{i<j}|y_{ij}|^2
 \prod_{\hat{i}<\hat{j}}|y_{\hat{i}\hat{j}}|^2
 \prod_{i,\hat{j}}|y_{i\hat{j}}|^{-2-2k}.
\end{eqnarray}
 Using the relation between quantum numbers
\begin{equation}
  \sum_a m_a = \frac{k}{2},~~
  \sum_a \bar{m}_a = -\frac{k}{2},~~
  \sum_a j_a +1 = \frac{k}{2}(2n+1),
\end{equation}
 we can indeed find that the above expression is independent of
 $z_a$ as expected from conformal invariance.
 If we put $z_1=0, z_2=1$ and $z_3=\infty$,
 it reduces to the formula (\ref{I_1}) in the previous section. 
 Ignoring the phase, we finally obtain the following expression:
\begin{eqnarray}
  A_+ &=&
 \delta^n(\tsum_i k_i)
 \delta(\tsum p_a)\delta(\tsum w_a-1)
\nonumber \\ &&
 \frac{\pi\mu_+^n\mu_-^{n+1}}{k\sin[(2n+1)\pi]}
 \frac{\Gamma(1+j_1-m_1)\Gamma(1+j_2-m_2)\Gamma(1+j_3-m_3)}
      {\Gamma(-j_1-\bar{m}_1)\Gamma(-j_2-\bar{m}_2)\Gamma(-j_3-\bar{m}_3)}
\nonumber \\ &&
 \prod_{s=1}^n
  \frac{\Gamma(-sk)\Gamma(1+2j_1-sk)\Gamma(1+2j_2-sk)\Gamma(1+2j_3-sk)}
       {\Gamma(1+sk)\Gamma(sk-2j_1)\Gamma(sk-2j_2)\Gamma(sk-2j_3)}.
\end{eqnarray}
 We see that the three-point function is manifestly symmetric
 in three vertices.
 Note that the string coupling was absorbed in the coupling
 $\mu_\pm$ by an appropriate shift of $\rho$.
 It can be restored using the standard scaling arguments of \cite{GM}.

\section{Conclusion}

   In this paper we analyzed the three-point functions of
 tachyons in sine-Liouville theory which is believed to be
 dual to the coset model $SL(2,{\bf R})/U(1)$.
 We checked that the sine-Liouville theory correctly
 reproduces the three-point functions in the coset model
 when the winding number is conserved.
 We also derived that the violation of total winding number
 is up to $\pm 1$ in three-point functions.
 All these are in agreement with the work of FZZ.
 We also obtained the expression for generic three-point
 functions when the winding number is non-conserving.

   In the case where the total winding number is $\pm 1$,
 it seems that the correlators in sine-Liouville theory do not
 fully reflect the $SL(2,{\bf R})$ group structure,
 since in general the Clebsch-Gordan coefficients for
 $SL(2,{\bf R})$ are expressed in terms of the generalized
 hypergeometric functions $\ghyp$.
 It would be interesting to study what the implication is.

   Obviously, the theories studied in this paper is closely
 related to the $SL(2,{\bf R})$ WZW model which describes
 a string on $AdS_3$.
 Our new result will give a new insight into the analysis
 of the processes involving a ``long string'' which has
 a macroscopic configuration of worldsheet winding around
 the boundary of $AdS_3$.
 This will be our future problem.

~

\noindent{\bf Note Added}

   After the first version of this paper was submitted, 
 the authors received some comments from V. Fateev that
 the same result was obtained in the unpublished work
 using essentially the same techniques.
 Their analysis used some integral identities 
 the spinless case of which can be found in \cite{BF}.
 Some comments about this duality can also be found in \cite{F}.

~

\noindent{\bf Acknowledgments}

   We are grateful to H. Kunitomo and Y. Satoh for useful
 discussions and comments.
 The works of the authors were supported in part by
 JSPS Research Fellowships for Young Scientists.

\newpage
\appendix

\section{New integral formulae}
   Let us summarize here our new formulae obtained in this paper.

   We calculated the integral of the form
\begin{eqnarray}
 I &=& \frac{1}{m!n!}\int\prod_{i=1}^m\left(
      d^2z_i z_i^\alpha\bar{z}_i^{\bar{\alpha}}
           (1-z_i)^\beta(1-\bar{z}_i)^{\bar{\beta}}\right) 
     \prod_{i<j}^m|z_{ij}|^2
    \nonumber \\ && ~~~~~~\cdot
   \prod_{\hat{i}=\hat{1}}^{n}\left(
      d^2z_{\hat{i}}
      z_{\hat{i}}^{\alpha'}\bar{z}_{\hat{i}}^{\bar{\alpha}'}
      (1-z_{\hat{i}})^{\beta'}(1-\bar{z}_{\hat{i}})^{\bar{\beta}'}\right) 
     \prod_{{\hat{i}}<{\hat{j}}}^n|z_{\hat{i}\hat{j}}|^2
     \prod_{i,\hat{j}}^{m,n}|z_{i\hat{j}}|^{2-2k}.
\end{eqnarray}
 This was found to be nonzero only when $m=n\pm 1$ or $m=n$.
 Introducing  $k'=k-2$ and the parameters
 $\gamma, \gamma', \bar{\gamma}$ and $\bar{\gamma}'$ satisfying
\begin{eqnarray*}
&&   \alpha+\beta+\gamma
 = \bar{\alpha}+\bar{\beta}+\bar{\gamma}
 = kn-m-n-1,~~ \\
&&   \alpha'+\beta'+\gamma'
 = \bar{\alpha}'+\bar{\beta}'+\bar{\gamma}'
 = km-m-n-1,~~
\end{eqnarray*}
 the integral becomes
\begin{eqnarray}
\lefteqn{(m=n+1)} \nonumber \\
  I &=& -\pi^{m+n}(-)^{\gamma-\bar{\gamma}}
        \frac{\Gamma(1+\alpha)\Gamma(1+\beta)\Gamma(1+\gamma)}
             {\Gamma(-\bar{\alpha})\Gamma(-\bar{\beta})\Gamma(-\bar{\gamma})}
 \nonumber \\ && \cdot
        \prod_{j=1}^n
     \frac{\Gamma(-jk')
           \Gamma(1+\alpha+\alpha'-jk')
           \Gamma(1+\beta+\beta'-jk')
           \Gamma(1+\gamma+\gamma'-jk')}
          {\Gamma(1+jk')
           \Gamma(jk'-\alpha-\alpha')
           \Gamma(jk'-\beta-\beta')
           \Gamma(jk'-\gamma-\gamma')},
  \nonumber \\
\lefteqn{(m=n)} \nonumber \\
 I&=& \pi^{2n-2}I_1
      \prod_{j=1}^{n-1}
      \frac{\Gamma(-jk')
            \Gamma(1+\alpha+\alpha'-jk')
            \Gamma(1+\beta +\beta' -jk')
            \Gamma(1+\gamma+\gamma'-jk')}
           {\Gamma(1+jk')
            \Gamma(jk'-\alpha-\alpha')
            \Gamma(jk'-\beta -\beta' )
            \Gamma(jk'-\gamma-\gamma')},
  \nonumber \\
 I_1&=&\int d^2z d^2w
     z^\alpha (1-z)^{\beta}
     \bar{z}^{\bar{\alpha}} (1-\bar{z})^{\bar{\beta}}
     w^{\alpha'} (1-w)^{\beta'}
     \bar{w}^{\bar{\alpha}'} (1-\bar{w})^{\bar{\beta}'}
     |z-w|^{-2-2nk'}.
\end{eqnarray}
 The four-dimensional integral $I_1$ can be expressed as
 $(a,b=1,2,3)$
 we can rewrite $I_1$ using certain double-contour integrals
 $c^{ab}$ and $p^{ab}$ as
\begin{equation}
  I_1 = 
   c^{12}[\alpha,\beta,\gamma]p^{12}[\bar{\alpha},\bar{\beta},\bar{\gamma}]
  +c^{21}[\alpha,\beta,\gamma]p^{21}[\bar{\alpha},\bar{\beta},\bar{\gamma}].
\end{equation}
 Using the function
 $G[^{a,b,c}_{~e,f}]\equiv
   \frac{\Gamma(a)\Gamma(b)\Gamma(c)}{\Gamma(e)\Gamma(f)}
  \ghyp(a,b,c;e,f;1)$, the integral
 $c^{12}$ is expressed as follows:
\begin{eqnarray}
 c^{12} &=&
  \frac{\Gamma(1+\alpha+\alpha'-nk')\Gamma(1+\beta+\beta'-nk')}
       {\Gamma(nk'-\gamma-\gamma')}
 G\IIxI{\alpha'+1,\beta+1,nk'-\gamma-\gamma'}
       {\alpha'-\gamma+1,\beta-\gamma'+1},
\end{eqnarray}
 and $c^{ab}$ are defined by the permutation of parameters.
 $p^{ab}$ are related to $c^{ab}$ via
\begin{equation}
  \IIxI{p^{12}}{p^{21}}
 = A_\beta\IIxI{c^{23}}{c^{32}}
 = A_\alpha^T\IIxI{c^{31}}{c^{13}},~~
 A_\alpha =
  \IIxII{\sinpi{\alpha}\sinpi{\alpha'}}{-\sinpi{\alpha}\sinpi{\alpha'-nk'}}
        {-\sinpi{\alpha'}\sinpi{\alpha-nk'}}{\sinpi{\alpha}\sinpi{\alpha'}}.
\end{equation}
 where $\sinpi{x}=\sin(\pi x)$.


\end{document}

%% file: psbox.tex
\def\temp{1.34}%
\let\tempp=\relax
\expandafter\ifx\csname psboxversion\endcsname\relax
\else
    \ifdim\temp cm>\psboxversion cm
    \else
      \let\temp=\psboxversion
      \let\tempp= 
    \fi
\fi
\tempp
\let\psboxversion=\temp
\catcode`\@=11
\def\psfortextures{
\def\PSspeci@l##1##2{%
\special{illustration ##1\space scaled ##2}%
}}%
\def\psfordvitops{
\def\PSspeci@l##1##2{%
\special{dvitops: import ##1\space \the\drawingwd \the\drawinght}%
}}%
\def\psfordvips{
\def\PSspeci@l##1##2{%
\d@my=0.1bp \d@mx=\drawingwd \divide\d@mx by\d@my
\includegraphics{##1\space}}}%
\def\psforoztex{
\def\PSspeci@l##1##2{%
\special{##1 \space
      ##2 1000 div dup scale
      \number-\psllx\space \number-\pslly\space translate
}}}%
\def\psfordvitps{
\def\psdimt@n@sp##1{\d@mx=##1\relax\edef\psn@sp{\number\d@mx}}
\def\PSspeci@l##1##2{%
\special{dvitps: Include0 "psfig.psr"}
\psdimt@n@sp{\drawingwd}
\special{dvitps: Literal "\psn@sp\space"}
\psdimt@n@sp{\drawinght}
\special{dvitps: Literal "\psn@sp\space"}
\psdimt@n@sp{\psllx bp}
\special{dvitps: Literal "\psn@sp\space"}
\psdimt@n@sp{\pslly bp}
\special{dvitps: Literal "\psn@sp\space"}
\psdimt@n@sp{\psurx bp}
\special{dvitps: Literal "\psn@sp\space"}
\psdimt@n@sp{\psury bp}
\special{dvitps: Literal "\psn@sp\space startTexFig\space"}
\special{dvitps: Include1 "##1"}
\special{dvitps: Literal "endTexFig\space"}
}}%
\def\psfordvialw{
\def\PSspeci@l##1##2{
\special{language "PostScript",
position = "bottom left",
literal "  \psllx\space \pslly\space translate
  ##2 1000 div dup scale
  -\psllx\space -\pslly\space translate",
include "##1"}
}}%
\def\psforptips{
\def\PSspeci@l##1##2{{
\d@mx=\psurx bp
\advance \d@mx by -\psllx bp
\divide \d@mx by 1000\multiply\d@mx by \xscale
\incm{\d@mx}
\let\tmpx\dimincm
\d@my=\psury bp
\advance \d@my by -\pslly bp
\divide \d@my by 1000\multiply\d@my by \xscale
\incm{\d@my}
\let\tmpy\dimincm
\d@mx=-\psllx bp
\divide \d@mx by 1000\multiply\d@mx by \xscale
\d@my=-\pslly bp
\divide \d@my by 1000\multiply\d@my by \xscale
\at(\d@mx;\d@my){\special{ps:##1 x=\tmpx, y=\tmpy}}
}}}%
\def\psonlyboxes{
\def\PSspeci@l##1##2{%
\at(0cm;0cm){\boxit{\vbox to\drawinght
  {\vss\hbox to\drawingwd{\at(0cm;0cm){\hbox{({\tt##1})}}\hss}}}}
}}%
\def\psloc@lerr#1{%
\let\savedPSspeci@l=\PSspeci@l%
\def\PSspeci@l##1##2{%
\at(0cm;0cm){\boxit{\vbox to\drawinght
  {\vss\hbox to\drawingwd{\at(0cm;0cm){\hbox{({\tt##1}) #1}}\hss}}}}
\let\PSspeci@l=\savedPSspeci@l
}}%
\newread\pst@mpin
\newdimen\drawinght\newdimen\drawingwd
\newdimen\psxoffset\newdimen\psyoffset
\newbox\drawingBox
\newcount\xscale \newcount\yscale \newdimen\pscm\pscm=1cm
\newdimen\d@mx \newdimen\d@my
\newdimen\pswdincr \newdimen\pshtincr
\let\ps@nnotation=\relax
{\catcode`\|=0 |catcode`|\=12 |catcode`|
|catcode`#=12 |catcode`*=14
|xdef|backslashother{\}*
|xdef|percentother{
|xdef|tildeother{~}*
|xdef|sharpother{#}*
}%
\def\R@moveMeaningHeader#1:->{}%
\def\uncatcode#1{%
\edef#1{\expandafter\R@moveMeaningHeader\meaning#1}}%
\def\execute#1{#1}
\def\psm@keother#1{\catcode`#112\relax}
\def\executeinspecs#1{%
\execute{\begingroup\let\do\psm@keother\dospecials\catcode`\^^M=9#1\endgroup}}%
\def\@mpty{}%
\def\matchexpin#1#2{
  \fi%
  \edef\tmpb{{#2}}%
  \expandafter\makem@tchtmp\tmpb%
  \edef\tmpa{#1}\edef\tmpb{#2}%
  \expandafter\expandafter\expandafter\m@tchtmp\expandafter\tmpa\tmpb\endm@tch%
  \if\match%
}%
\def\matchin#1#2{%
  \fi%
  \makem@tchtmp{#2}%
  \m@tchtmp#1#2\endm@tch%
  \if\match%
}%
\def\makem@tchtmp#1{\def\m@tchtmp##1#1##2\endm@tch{%
  \def\tmpa{##1}\def\tmpb{##2}\let\m@tchtmp=\relax%
  \ifx\tmpb\@mpty\def\match{YN}%
  \else\def\match{YY}\fi%
}}%
\def\incm#1{{\psxoffset=1cm\d@my=#1
 \d@mx=\d@my
  \divide\d@mx by \psxoffset
  \xdef\dimincm{\number\d@mx.}
  \advance\d@my by -\number\d@mx cm
  \multiply\d@my by 100
 \d@mx=\d@my
  \divide\d@mx by \psxoffset
  \edef\dimincm{\dimincm\number\d@mx}
  \advance\d@my by -\number\d@mx cm
  \multiply\d@my by 100
 \d@mx=\d@my
  \divide\d@mx by \psxoffset
  \xdef\dimincm{\dimincm\number\d@mx}
}}%
\newif\ifNotB@undingBox
\newhelp\PShelp{Proceed: you'll have a 5cm square blank box instead of
your graphics (Jean Orloff).}%
\def\s@tsize#1 #2 #3 #4\@ndsize{
  \def\psllx{#1}\def\pslly{#2}%
  \def\psurx{#3}\def\psury{#4}
  \ifx\psurx\@mpty\NotB@undingBoxtrue
  \else
    \drawinght=#4bp\advance\drawinght by-#2bp
    \drawingwd=#3bp\advance\drawingwd by-#1bp
  \fi
  }%
\def\sc@nBBline#1:#2\@ndBBline{\edef\p@rameter{#1}\edef\v@lue{#2}}%
\def\g@bblefirstblank#1#2:{\ifx#1 \else#1\fi#2}%
{\catcode`\%=12
\xdef\B@undingBox{
\def\ReadPSize#1{
 \readfilename#1\relax
 \let\PSfilename=\lastreadfilename
 \openin\pst@mpin=#1\relax
 \ifeof\pst@mpin \errhelp=\PShelp
   \errmessage{I haven't found your postscript file (\PSfilename)}%
   \psloc@lerr{was not found}%
   \s@tsize 0 0 142 142\@ndsize
   \closein\pst@mpin
 \else
   \if\matchexpin{\GlobalInputList}{, \lastreadfilename}%
   \else\xdef\GlobalInputList{\GlobalInputList, \lastreadfilename}%
     \immediate\write\psbj@inaux{\lastreadfilename,}%
   \fi%
   \loop
     \executeinspecs{\catcode`\ =10\global\read\pst@mpin to\n@xtline}%
     \ifeof\pst@mpin
       \errhelp=\PShelp
       \errmessage{(\PSfilename) is not an Encapsulated PostScript File:
           I could not find any \B@undingBox: line.}%
       \edef\v@lue{0 0 142 142:}%
       \psloc@lerr{is not an EPSFile}%
       \NotB@undingBoxfalse
     \else
       \expandafter\sc@nBBline\n@xtline:\@ndBBline
       \ifx\p@rameter\B@undingBox\NotB@undingBoxfalse
         \edef\t@mp{%
           \expandafter\g@bblefirstblank\v@lue\space\space\space}%
         \expandafter\s@tsize\t@mp\@ndsize
       \else\NotB@undingBoxtrue
       \fi
     \fi
   \ifNotB@undingBox\repeat
   \closein\pst@mpin
 \fi
\message{#1}%
}%
\def\psboxto(#1;#2)#3{\vbox{%
   \ReadPSize{#3}%
   \advance\pswdincr by \drawingwd
   \advance\pshtincr by \drawinght
   \divide\pswdincr by 1000
   \divide\pshtincr by 1000
   \d@mx=#1
   \ifdim\d@mx=0pt\xscale=1000
         \else \xscale=\d@mx \divide \xscale by \pswdincr\fi
   \d@my=#2
   \ifdim\d@my=0pt\yscale=1000
         \else \yscale=\d@my \divide \yscale by \pshtincr\fi
   \ifnum\yscale=1000
         \else\ifnum\xscale=1000\xscale=\yscale
                    \else\ifnum\yscale<\xscale\xscale=\yscale\fi
              \fi
   \fi
   \divide\drawingwd by1000 \multiply\drawingwd by\xscale
   \divide\drawinght by1000 \multiply\drawinght by\xscale
   \divide\psxoffset by1000 \multiply\psxoffset by\xscale
   \divide\psyoffset by1000 \multiply\psyoffset by\xscale
   \global\divide\pscm by 1000
   \global\multiply\pscm by\xscale
   \multiply\pswdincr by\xscale \multiply\pshtincr by\xscale
   \ifdim\d@mx=0pt\d@mx=\pswdincr\fi
   \ifdim\d@my=0pt\d@my=\pshtincr\fi
   \message{scaled \the\xscale}%
 \hbox to\d@mx{\hss\vbox to\d@my{\vss
   \global\setbox\drawingBox=\hbox to 0pt{\kern\psxoffset\vbox to 0pt{%
      \kern-\psyoffset
      \PSspeci@l{\PSfilename}{\the\xscale}%
      \vss}\hss\ps@nnotation}%
   \global\wd\drawingBox=\the\pswdincr
   \global\ht\drawingBox=\the\pshtincr
   \global\drawingwd=\pswdincr
   \global\drawinght=\pshtincr
   \baselineskip=0pt
   \copy\drawingBox
 \vss}\hss}%
  \global\psxoffset=0pt
  \global\psyoffset=0pt
  \global\pswdincr=0pt
  \global\pshtincr=0pt 
  \global\pscm=1cm 
}}%
\def\psboxscaled#1#2{\vbox{%
  \ReadPSize{#2}%
  \xscale=#1
  \message{scaled \the\xscale}%
  \divide\pswdincr by 1000 \multiply\pswdincr by \xscale
  \divide\pshtincr by 1000 \multiply\pshtincr by \xscale
  \divide\psxoffset by1000 \multiply\psxoffset by\xscale
  \divide\psyoffset by1000 \multiply\psyoffset by\xscale
  \divide\drawingwd by1000 \multiply\drawingwd by\xscale
  \divide\drawinght by1000 \multiply\drawinght by\xscale
  \global\divide\pscm by 1000
  \global\multiply\pscm by\xscale
  \global\setbox\drawingBox=\hbox to 0pt{\kern\psxoffset\vbox to 0pt{%
     \kern-\psyoffset
     \PSspeci@l{\PSfilename}{\the\xscale}%
     \vss}\hss\ps@nnotation}%
  \advance\pswdincr by \drawingwd
  \advance\pshtincr by \drawinght
  \global\wd\drawingBox=\the\pswdincr
  \global\ht\drawingBox=\the\pshtincr
  \global\drawingwd=\pswdincr
  \global\drawinght=\pshtincr
  \baselineskip=0pt
  \copy\drawingBox
  \global\psxoffset=0pt
  \global\psyoffset=0pt
  \global\pswdincr=0pt
  \global\pshtincr=0pt 
  \global\pscm=1cm
}}%
\def\psbox#1{\psboxscaled{1000}{#1}}%
\newif\ifn@teof\n@teoftrue
\newif\ifc@ntrolline
\newif\ifmatch
\newread\j@insplitin
\newwrite\j@insplitout
\newwrite\psbj@inaux
\immediate\openout\psbj@inaux=psbjoin.aux
\immediate\write\psbj@inaux{\string\joinfiles}%
\immediate\write\psbj@inaux{\jobname,}%
\def\toother#1{\ifcat\relax#1\else\expandafter%
  \toother@ux\meaning#1\endtoother@ux\fi}%
\def\toother@ux#1 #2#3\endtoother@ux{\def\tmp{#3}%
  \ifx\tmp\@mpty\def\tmp{#2}\let\next=\relax%
  \else\def\next{\toother@ux#2#3\endtoother@ux}\fi%
\next}%
\let\readfilenamehook=\relax
\def\re@d{\expandafter\re@daux}
\def\re@daux{\futurelet\nextchar\stopre@dtest}%
\def\re@dnext{\xdef\lastreadfilename{\lastreadfilename\nextchar}%
  \afterassignment\re@d\let\nextchar}%
\def\stopre@d{\egroup\readfilenamehook}%
\def\stopre@dtest{%
  \ifcat\nextchar\relax\let\nextread\stopre@d
  \else
    \ifcat\nextchar\space\def\nextread{%
      \afterassignment\stopre@d\chardef\nextchar=`}%
    \else\let\nextread=\re@dnext
      \toother\nextchar
      \edef\nextchar{\tmp}%
    \fi
  \fi\nextread}%
\def\readfilename{\bgroup%
  \let\\=\backslashother \let\%=\percentother \let\~=\tildeother
  \let\#=\sharpother \xdef\lastreadfilename{}%
  \re@d}%
\xdef\GlobalInputList{\jobname}%
\def\psnewinput{%
  \def\readfilenamehook{
    \if\matchexpin{\GlobalInputList}{, \lastreadfilename}%
    \else\xdef\GlobalInputList{\GlobalInputList, \lastreadfilename}%
      \immediate\write\psbj@inaux{\lastreadfilename,}%
    \fi%
    \ps@ldinput\lastreadfilename\relax%
    \let\readfilenamehook=\relax%
  }\readfilename%
}%
\expandafter\ifx\csname @@input\endcsname\relax    
  \immediate\let\ps@ldinput=\input\def\input{\psnewinput}%
\else
  \immediate\let\ps@ldinput=\@@input
  \def\@@input{\psnewinput}%
\fi%
\def\nowarnopenout{%
 \def\warnopenout##1##2{%
   \readfilename##2\relax
   \message{\lastreadfilename}%
   \immediate\openout##1=\lastreadfilename\relax}}%
\def\warnopenout#1#2{%
 \readfilename#2\relax
 \def\t@mp{TrashMe,psbjoin.aux,psbjoint.tex,}\uncatcode\t@mp
 \if\matchexpin{\t@mp}{\lastreadfilename,}%
 \else
   \immediate\openin\pst@mpin=\lastreadfilename\relax
   \ifeof\pst@mpin
     \else
     \errhelp{If the content of this file is so precious to you, abort (ie
press x or e) and rename it before retrying.}%
     \errmessage{I'm just about to replace your file named \lastreadfilename}%
   \fi
   \immediate\closein\pst@mpin
 \fi
 \message{\lastreadfilename}%
 \immediate\openout#1=\lastreadfilename\relax}%
{\catcode`\%=12\catcode`\*=14
\gdef\splitfile#1{*
 \readfilename#1\relax
 \immediate\openin\j@insplitin=\lastreadfilename\relax
 \ifeof\j@insplitin
   \message{! I couldn't find and split \lastreadfilename!}*
 \else
   \immediate\openout\j@insplitout=TrashMe
   \message{< Splitting \lastreadfilename\space into}*
   \loop
     \ifeof\j@insplitin
       \immediate\closein\j@insplitin\n@teoffalse
     \else
       \n@teoftrue
       \executeinspecs{\global\read\j@insplitin to\spl@tinline\expandafter
         \ch@ckbeginnewfile\spl@tinline
       \ifc@ntrolline
       \else
         \toks0=\expandafter{\spl@tinline}*
         \immediate\write\j@insplitout{\the\toks0}*
       \fi
     \fi
   \ifn@teof\repeat
   \immediate\closeout\j@insplitout
 \fi\message{>}*
}*
\gdef\ch@ckbeginnewfile#1
 \def\t@mp{#1}*
 \ifx\@mpty\t@mp
   \def\t@mp{#3}*
   \ifx\@mpty\t@mp
     \global\c@ntrollinefalse
   \else
     \immediate\closeout\j@insplitout
     \warnopenout\j@insplitout{#2}*
     \global\c@ntrollinetrue
   \fi
 \else
   \global\c@ntrollinefalse
 \fi}*
\gdef\joinfiles#1\into#2{*
 \message{< Joining following files into}*
 \warnopenout\j@insplitout{#2}*
 \message{:}*
 {*
 \edef\w@##1{\immediate\write\j@insplitout{##1}}*
\w@{
\w@{
\w@{
\w@{
\w@{
\w@{
\w@{
\w@{
\w@{
\w@{
\w@{\string\input\space psbox.tex}*
\w@{\string\splitfile{\string\jobname}}*
\w@{\string\let\string\autojoin=\string\relax}*
}*
 \expandafter\tre@tfilelist#1, \endtre@t
 \immediate\closeout\j@insplitout
 \message{>}*
}*
\gdef\tre@tfilelist#1, #2\endtre@t{*
 \readfilename#1\relax
 \ifx\@mpty\lastreadfilename
 \else
   \immediate\openin\j@insplitin=\lastreadfilename\relax
   \ifeof\j@insplitin
     \errmessage{I couldn't find file \lastreadfilename}*
   \else
     \message{\lastreadfilename}*
     \immediate\write\j@insplitout{
     \executeinspecs{\global\read\j@insplitin to\oldj@ininline}*
     \loop
       \ifeof\j@insplitin\immediate\closein\j@insplitin\n@teoffalse
       \else\n@teoftrue
         \executeinspecs{\global\read\j@insplitin to\j@ininline}*
         \toks0=\expandafter{\oldj@ininline}*
         \let\oldj@ininline=\j@ininline
         \immediate\write\j@insplitout{\the\toks0}*
       \fi
     \ifn@teof
     \repeat
   \immediate\closein\j@insplitin
   \fi
   \tre@tfilelist#2, \endtre@t
 \fi}*
}%
\def\autojoin{%
 \immediate\write\psbj@inaux{\string\into{psbjoint.tex}}%
 \immediate\closeout\psbj@inaux
 \expandafter\joinfiles\GlobalInputList\into{psbjoint.tex}%
}%
\def\centinsert#1{\midinsert\line{\hss#1\hss}\endinsert}%
\def\psannotate#1#2{\vbox{%
  \def\ps@nnotation{#2\global\let\ps@nnotation=\relax}#1}}%
\def\pscaption#1#2{\vbox{%
   \setbox\drawingBox=#1
   \copy\drawingBox
   \vskip\baselineskip
   \vbox{\hsize=\wd\drawingBox\setbox0=\hbox{#2}%
     \ifdim\wd0>\hsize
       \noindent\unhbox0\tolerance=5000
    \else\centerline{\box0}%
    \fi
}}}%
\def\at(#1;#2)#3{\setbox0=\hbox{#3}\ht0=0pt\dp0=0pt
  \rlap{\kern#1\vbox to0pt{\kern-#2\box0\vss}}}%
\newdimen\gridht \newdimen\gridwd
\def\gridfill(#1;#2){%
  \setbox0=\hbox to 1\pscm
  {\vrule height1\pscm width.4pt\leaders\hrule\hfill}%
  \gridht=#1
  \divide\gridht by \ht0
  \multiply\gridht by \ht0
  \gridwd=#2
  \divide\gridwd by \wd0
  \multiply\gridwd by \wd0
  \advance \gridwd by \wd0
  \vbox to \gridht{\leaders\hbox to\gridwd{\leaders\box0\hfill}\vfill}}%
\def\fillinggrid{\at(0cm;0cm){\vbox{%
  \gridfill(\drawinght;\drawingwd)}}}%
\def\textleftof#1:{%
  \setbox1=#1
  \setbox0=\vbox\bgroup
    \advance\hsize by -\wd1 \advance\hsize by -2em}%
\def\textrightof#1:{%
  \setbox0=#1
  \setbox1=\vbox\bgroup
    \advance\hsize by -\wd0 \advance\hsize by -2em}%
\def\endtext{%
  \egroup
  \hbox to \hsize{\valign{\vfil##\vfil\cr%
\box0\cr%
\noalign{\hss}\box1\cr}}}%
\def\frameit#1#2#3{\hbox{\vrule width#1\vbox{%
  \hrule height#1\vskip#2\hbox{\hskip#2\vbox{#3}\hskip#2}%
        \vskip#2\hrule height#1}\vrule width#1}}%
\def\boxit#1{\frameit{0.4pt}{0pt}{#1}}%
\catcode`\@=12 
 \psfordvips   